\documentclass[prb,twocolumn,showpacs,amsmath,amssymb,superscriptaddress]{revtex4-1}
\usepackage{graphicx}% Include figure files
\usepackage{dcolumn}% Align table columns on decimal point
\usepackage{bm}
\usepackage{bbm}

\newcommand{\nc}{\newcommand}

\nc{\be}{\begin{equation}} \nc{\ee}{\end{equation}}
\nc{\bea}{\begin{eqnarray}} \nc{\eea}{\end{eqnarray}}
\nc{\bean}{\begin{eqnarray*}} \nc{\eean}{\end{eqnarray*}}

\begin{document}

\title{%Spin dynamics in the presence of the general spin-orbit coupling with the particle-particle interactions
Unified theory of spin-dynamics in  a two dimensional electron gase with arbitrary spin-orbit coping strength at finite temperature}
\author{Xin Liu}
\affiliation{ Department of Physics, Texas A\&M University, College
Station, TX 77843-4242, USA}
\author{Jairo Sinova}
\affiliation{ Department of Physics, Texas A\&M University, College
Station, TX 77843-4242, USA}
\affiliation{Institute of Physics ASCR, Cukrovarnick\'a 10, 162 53 Praha 6, Czech
Republic }
\date{\today}

\begin{abstract}
We study the spin dynamics in the presence of impurity and electron-electron (e-e) scattering in a 
III-V semiconductor quantum well with arbitrary spin-orbit coupling (SOC) strength and symmetry at finite temperature. 
%Starting from the non-equilibrium Green's function formalism, 
We derive the coupled spin-charge dynamic equations %at finite temperature
 in the presence of inelastic  scattering 
and provide a new formalism that describes the spin relaxation and dynamics 
in both the weak and the strong SOC regime in a unified way.
 In the weak SOC regime, as expected, our theory reproduces all previous zero-temperature results, most of which 
 have focused on impurity-scattering induced spin-charge dynamics. In the regime where the strength of the  Rashba and 
 linear Dresselhaus  SOC match, known as the SU(2) symmetry point, experiments have observed the spin-helix mode with 
 a large spin-lifetime whose unexplained non-monotonic temperature dependence peaks at  around $75$ K. 
 As a key test of our theory, we are able to naturally explain { quantitatively} this non-monotonic dependence 
 and show that it arises as a competition between the 
 Dyakonov-Perel mechanism, suppressed at the SU(2) point, and the Elliott-Yafet mechanism. 
 In the strong SOC regime, we show that our theory directly reproduces 
 the only previous known analytical result at the SU(2) symmetry point
 in the ballistic regime.  It also explains, as we have shown previously, the rise of damped oscillating dynamics   
when the electron scattering time is larger than half of the spin precession time due to the SOC.
Hence, we provide a unified theory of the spin-dynamics in two dimensional electron gases in the full phase diagram
experimentally accessible. 
%which should prove useful in the modeling of future devices.
% 
% our theory shows that when the system is near the SU(2) symmetry point~\cite{Bernevig:2006_a}, because the spin relaxation due to DP mechanism is suppressed dramatically \cite{Stanescu:2007_a}, the spin relaxation is dominated by the Elliott-Yafet (EY) mechanism in a wide temperature regime. The non-monotonic temperature dependence of enhanced-lifetime of spin helix mode \cite{Koralek:2009_a} is due to the competition between the DP and EY mechanisms. In the strong SOC regime, the our theory is consistent to the previous theoretical results \cite{Bernevig:2008_a,LiuXin:2011_a} at zero temperature.
\end{abstract}

\maketitle

\section{Introduction}

The spin degree freedom of electrons and the control of its dynamics by electric means 
has played an increasingly  important role in condensed matter research and spintronics device applications
over the past decades. This electric manipulation is achieved via the SOC, which has been a key factor in new
 spintronic devices as well as the emergence of new fields 
such as spin Hall effect,\cite{Hirsch:1999_a, Murakami:2003_a, Sinova:2004_a} spin Coulomb drag,~\cite{Damico:2000_a,Damico:2003_a,Weber:2005_a} and  topological insulators.~\cite{Hasan:2010_a,Qi:2011_a}

These fast developing SOC-based fields highlight the need to fully understand the basic spin-charge transport dynamics in the 
full phase diagram experimentally accessible, both at zero and finite temperature. 
%However spin containing two internal states provides the great challenge for the theorists in dealing with the multi-bands transport and dynamic problems such as spin relaxation \cite{Mishchenko:2004_a, Burkov:2004_a}. 
The spin relaxation and dynamics in the presence of weak SOC %(relative to disorder scattering) 
is a mature topic of research, 
and a theoretical perturbative treatment on the SOC 
gives an accurate description of the spin dynamics at zero temperature.~\cite{Burkov:2004_a, Mishchenko:2004_a, Bernevig:2006_a, Stanescu:2007_a} 
In the opposite limit of strong SOC, defined as $\Omega_{so}\tau >1$ where $\Omega_{so}$ is the spin precession frequency due 
to the SOCs and $\tau$ is the momentum scattering time,
only a 
few theoretical works~\cite{Bernevig:2006_a,Bernevig:2008_a}  were able to derive 
analytically the spin dynamics  at one specific point in the phase diagram; specifically when
the Rashba SOC strength is equal to the linear Dresselhaus SOC strength,
where a novel  spin dynamic mode, the spin-helix mode (SHM), with large spin-lifetimes 
has been observed experimentally.~\cite{Weber:2005_a} 
One theory work  was able to analytically derive the spin-dynamics modes at zero temperature
and successfully explain the different damped oscillatory modes observed in this strong SOC regime.\cite{LiuXin:2011_a} 
However, until now, analytical theories of the spin dynamics in 2DEGs 
have primarily focused on zero temperature.~\cite{Mishchenko:2004_a,Bernevig:2006_a,Bernevig:2008_a,Koralek:2009_a} 
The rich physics observed in the experiments as a function of temperature, which 
depict the transition from the strong to weak SOC 
and shows a peak of the spin lifetime of the SHE mode at $\sim 75$ K,
has remained  unexplained up to now.~\cite{Koralek:2009_a}

\begin{figure}
\centering
%\begin{tabular}{l}
\includegraphics[width=0.9\columnwidth]{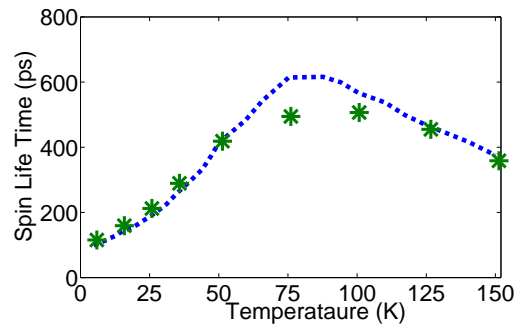}
%\end{tabular}
\vspace {-0.5 cm}
\caption{Spin life time of the dominant spin-helix mode vs. temperature,  extracted from the Ref.~\onlinecite{Koralek:2009_a} (Green stars). The blue dash line is our theoretical result (see Sec.~\ref{T-dep-of-spin-relaxation}).}  %by substituting $\tau$ in Fig.~\ref{life-T-1} to our spin dynamic equation.}
\label{life-T}
\vspace {-0.5 cm}
\end{figure}
Here we develop a microscopic approach valid at finite temperature and for any strength of SOC 
within the same derived expression. 
To show the validity of our method, we focus on the spin dynamics in the n-type III-V quantum well which has been 
recently well studied experimentally.\cite{Brand:2002_a,Weber:2005_a,Leyland:2007_a,Leyland:2007_b,Koralek:2009_a}
Besides reproducing all known limits and previously derived results at zero temperature,\cite{Burkov:2004_a, Mishchenko:2004_a, Stanescu:2007_a, Bernevig:2008_a, LiuXin:2011_a} it succeeds 
dramatically in describing quantitatively the non-monotonic behavior of the spin lifetime of the SHE mode as shown in 
Fig.~\ref{life-T}. Previously it was speculated that this non-monotonic behavior could be attributed to the increase in importance 
of the cubic Dresselhaus SOC, which decreases the SHM lifetime. However, a simple analysis (see Sec. III.A) shows that
in the range from zero to 100 K this term does not change significantly and, therefore, it fails to account for this behavior.
Our theory shows directly that this non-monotonic behavior arises from the  competition between the 
 Dyakonov-Perel (DP) mechanism, suppressed at the SU(2) point, and the Elliott-Yafet (EY) mechanism.

Our microscopic approach is based on the non-equilibrium Green's function formalism~\cite{Rammer:1986_a,Mishchenko:2004_a} 
which allows us to fully understand the temperature dependence of the spin relaxation in 
the presence of both disorder and e-e inelastic scattering.  
The e-e interaction is shown to dominate the momentum scattering time above a certain finite temperature 
(35 K in  the relevant experiments)
and controls the transition between strong and weak SOC regime.\cite{Leyland:2007_b} 
The e-e scattering time in the 2DEG system has been theoretically discussed previously\cite{Zheng:1996_a} and,
even though it does not affect the charge transport due to net momentum conservation of the system (see appendix \ref{e-e-c}), 
the spin-current and spin decay can be affected strongly by it, e.g. 
spin Coulomb drag.\cite{Damico:2000_a,Damico:2003_a,Weber:2005_a} 
By incorporating both spin independent scattering and 
SOC scattering potential, both DP and EY mechanisms are considered in the frame of the quantum kinetic equation.

Our paper is organized as follows. In Sec. \ref{QKE} we introduce the quantum kinetic equation (QKE) based on the non-equilibrium Green's function formalism in the presence of the three types of SOC present in  \uppercase\expandafter{\romannumeral3}-\uppercase\expandafter{\romannumeral5} based 2DEG. Within Sec. \ref{e-e-collision}, because we want to consider finite temperature, we evaluate the contribution of the  e-e interaction in the collision integral of the QKE and show that it dominates the momentum scattering time 
 for temperatures above $35$ K. 
 In Sec. \ref{QKE-thermal-average} we generalize the spin-charge kinetic equation from the  zero temperature limit \cite{Mishchenko:2004_a} to the finite temperature in the presence of  inelastic scattering. Within Sec. \ref{general-QKE}  we present our
 procedure which allows us to have a theory valid in both the weak and the strong SOC regime, with Eq. (\ref{GF-12}) and (\ref{GF-13}) being 
 the pivotal results that allows us to evaluate the full phase diagram of the spin-charge dynamics. 

In Sec. \ref{PSH}, we focus on the temperature dependence of lifetime of the dominant SHM near the SU(2) symmetry point,
were we show that the thermal average of the SOCs strength is almost unchanged from $0\rm{K}$ to $100\rm{K}$.
This discards the increase of the cubic Dresselhaus strength as source of the non-monotonic behavior of the 
spin-lifetime as mentioned above.\cite{Koralek:2009_a} In this section
we evaluate the EY mechanism and show quantitatively that the non-monotonic temperature dependence of the enhanced-lifetime of the SHM is the result of the competition of the EY and DP spin relaxation mechanism.
 Finally  in Sec. \ref{SSOC}, we reproduce the spin relaxation eigen-modes in the presence of the different SOCs which confirm that our method is also valid in the strong SOC regime.

 %Several recently experiments \cite{Brand:2002_a, Weber:2005_a, Leyland:2007_a} has observed the n-type GaAs/AlGaAs 2DEG undergoes the transition from the weak to the strong SOC regime when decreasing temperature from the room temperature to the order of $1$K. This transition seems to be dominated by e-e interaction \cite{Leyland:2007_b}. 
 %It was widely assumed that the spin current is unaffected by the e-e interaction like the charge current because spin and charge share the same carrier. As a result, according to the generalized Einstein relation of spin current and spin diffusive constant, the e-e interaction also should not affect the spin dynamics. However, I. D¡¯Amico and G. Vignale \cite{Damico:2000_a} first pointed out theoretically that the assumption is not valid and the e-e interaction provides the additional friction-like force to the spin current which is named as spin Coulomb drag \cite{Damico:2000_a}. Lately this theory was confirmed experimentally \cite{Weber:2005_a} by the nonuniform spin polarization dynamics. On the other hand, M. A. Brand \textit{et al}. \cite{Brand:2002_a}, found that they have to put the e-e interaction in their numerical simulation to fit their temperature behavior of the uniform spin polarization dynamics they observed experimentally. These facts demonstrate that the spin dynamics and the spin current do feel the e-e interaction in contrast to the charge current. Therefore it is important to develop an unified theory to understand why e-e interaction plays so different roles in the spin dynamics and charge transport.

\section{The quantum kinetic equation for the 2DEG with the general SOC}\label{QKE}

The non-equilibrium spin polarization can be described by the non-equilibrium Green's function formalism, 
also known as Keldysh formalism.~\cite{Rammer:1986_a} This formalism was used
in the noninteracting system with short range disorder and in the presence of weak Rashba SOC~\cite{Mishchenko:2004_a} or the equal magnitude of Rashba and linear Dresselhaus SOCs with zero cubic Dresselhaus SOC.~\cite{Bernevig:2008_a} However, when considering the temperature dependence of the spin life time, we have to generalize this method to the interacting system with 
a general SOCs. Here, we focus on the 2DEG in a quantum well such as GaAs/AlGaAs.\cite{Brand:2002_a,Weber:2005_a,Weber:2007_a,Koralek:2009_a}

In the 2-D semiconductor heterostructures, the Hamiltonian takes the form
\begin{eqnarray}\label{ha}
H=\frac{k^2}{2m}+\mathbf{b(k)} \cdot \bm{\hat{\sigma}},
\end{eqnarray}
where $\mathbf{b(k)}$ is the effective magnetic field and contains three types of SOCs, namely the linear Rashba \cite{Rashba:1960_a,Bychkov:1984_a} SOC and the linear and cubic Dresselhaus \cite{Dresselhaus:1955_a} SOCs which take the form
\begin{eqnarray}\label{soi}
\mathbf{b}^R(\mathbf{k})&=&\alpha (-k_y,k_x),\\
\mathbf{b}^{D_1} (\mathbf{k})&=&\beta_1 (k_y,k_x),\\
\mathbf{b}^{D_3}(\mathbf{k})&=&-2\beta_3\cos2\theta(-k_y,k_x),
\end{eqnarray}
where $k_F$ is the Fermi wave vector. Here we take $\theta$ as the angle between the wave vector $\mathbf{k}$ and the $[110]$ direction which we take to be the local $x$-axis in our coordinates.
 The above SOCs split the spin-degenerate bands and dominate the spin dynamics in the 2DEG. The corresponding SOC Hamiltonian takes the form:
\begin{eqnarray}\label{hso}
\hat{H}_{so}&=&\lambda_1 k_x\sigma_y+\lambda_2k_y\sigma_x=b_y\sigma_y+b_x\sigma_x,
\end{eqnarray}
where
$\lambda_1=\alpha+\beta_1-2\beta_3\cos2\theta$, $\lambda_2=\beta_1-\alpha+2\beta_3\cos2\theta$.
The retarded (advanced) Green's function of the Hamiltonian in Eq.~(\ref{ha}) takes the form
\begin{eqnarray}\label{Green-1}
G^{R(A)}(E,k)=\frac{(E-\frac{k^2}{2m})\sigma_0+\bm{b}(\bm{k})\cdot \bm{\sigma}}{(E-\frac{k^2}{2m}\pm i\delta)^2-b_{so}^2}.
\end{eqnarray}

 The nonequilibrium state of the system can be described by introducing the contour-ordering Green's function in the Keldysh space as
\begin{eqnarray}\label{qb-2}
\hat{G}(1,2)=\left(\begin{array}{cc} G^R(1,2) & G^K(1,2)\\ 0 & G^A(1,2)
\end{array} \right), \end{eqnarray}
where $1$ and $2$ stand for the condensed notation $1=(\bm{x}_1,s_{z1},t_1)$.

\begin{figure}
\centering
\begin{tabular}{l}
\includegraphics[width=1.0\columnwidth]{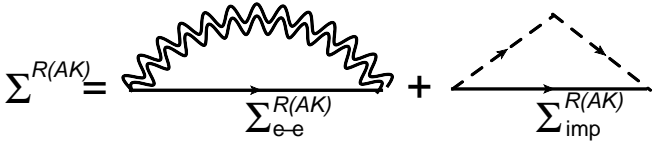}
\end{tabular}
\caption{The two self energy we consider in this work. The double wiggle in the first diagram is the effective e-e interaction in the random phase approximation (RPA). The dashed line in the second diagram represents the impurity scattering.}
\label{sigma-1}
\end{figure}

In the presence of these general SOCs, the kinetic part of the quantum Boltzmann equation has the form \cite{Rammer:2007_a,Mishchenko:2004_a}
\begin{eqnarray}\label{qbe-1}
&&\partial_T G^K+\frac{1}{2}\{\hat{\mathbf{V}},\cdot \mathbf{\nabla_R}G^K\}+i[\mathbf{b}(\mathbf{k}) \cdot \bm{\sigma},G^K]\nonumber \\
&=&-i\left[(\Sigma^R G^K-G^{K}\Sigma^A)-(G^R\Sigma^K-\Sigma^KG^A)\right], \nonumber \\
\end{eqnarray}
where $\hat{V}=\frac{\partial H}{\partial k}$, $[\cdots,\cdots]$ and $\{\cdots,\cdots\}$ stand for commutation relation and anti-commutation relation respectively, and $\Sigma^R$, $\Sigma^A$ and $\Sigma^K$ are the retarded, advanced and Keldysh self energy and the corresponding Feynman diagrams are shown in Fig.~\ref{sigma-1}. The self energy of the e-e interaction have the more complicated forms \cite{Schutt:2011_a}
\begin{eqnarray}\label{self-energy}
 &&\Sigma^{R(A)}_{ee}=\underline{G}^K\circ D^{R(A)}+\underline{G}^{R(A)}\circ D^K,\nonumber \\
 &&\Sigma^{K}_{ee}=(\underline{G}^R-\underline{G}^A)\circ (D^R-D^A)+\underline{G}^K \circ D^K,
 \end{eqnarray}
 where the symbol $\circ$ denotes integration over all internal energies and momenta, $D^{R(A,K)}$ are the full dressed propagator of Coulomb interaction \cite{Schutt:2011_a} and $\underline{G}^{R(A,K)}$ is the electron Green's function in the self energy diagram Fig.~\ref{sigma-1} and its energy and momentum are denoted as $E'$ and $\rm{k}'$ respectively. The underline is to emphasize its difference to $G^{R(A,K)}$ which is the electron Green's function out of the self energy; i.e. the bare self-energy.
  In the limit $t_1=t_2$ and $\rm{x}_1=\rm{x}_2$, $G^K(t_1=t_2,\rm{x}_1=\rm{x}_2)=1-2\hat{n}(x,t)$ where $\hat{n}(x,t)=\psi^{\dagger}(x,t)\psi(x,t)$ is the electron density operator.

\subsection{Collision integral of e-e interaction}
\label{e-e-collision}

The treatment of the collision integral contribution from the impurity scattering has been well established.~\cite{Mishchenko:2004_a} Therefore  we will focus on the e-e interaction in the collision integral
which may not be as  familiar to the readers and which, as shown below, can dominate the 
electron's momentum relaxation time at finite temperatures, even if 
it has no direct effect on the total momentum of the system. The first e-e scattering term has the form
\begin{eqnarray}\label{self-energy-1}
&&\Sigma^R_{ee} G^K-G^K \Sigma^A_{ee} \nonumber \\
&=&\left(\underline{G}^K\circ (D^R-D^A)+(\underline{G}^R-\underline{G}^A)\circ D^K\right)G^K.
\end{eqnarray}
The Keldysh Green's function can be written as $G^K=G^K_0+\delta G^K$ where $G^K_0 (\delta G^K)$ is the Keldysh Green's function of the equilibrium (non-equilibrium) part. The terms containing the first order of $\delta G^K$ and $\delta\underline{G}^K$ in Eq.~(\ref{self-energy-1}) have the form
\begin{eqnarray}\label{self-energy-1-1}
&&\left(\underline{G}^K_0\circ (D^R-D^A)+(\underline{G}^R-\underline{G}^A)\circ D^K\right)\delta G^K\nonumber \\ &&+ \delta \underline{G}^K_0\circ (D^R-D^A) G^K_0\nonumber \\
&=& \frac{i}{\tau_{ee}} \delta G^K+\delta \underline{G}^K\circ (D^R-D^A) G^K_0\, ,
\end{eqnarray}
where~\cite{Schutt:2011_a}
\begin{eqnarray}\label{self-energy-1-2}
\frac{1}{\tau_{ee}}&=&\left(\underline{G}^K_0\circ (D^R-D^A)+(\underline{G}^R-\underline{G}^A)\circ D^K\right)\nonumber \\
&=&\int \frac{dE'}{2\pi} \frac{d^2k'}{(2\pi)^2}(D^R-D^A)(\underline{G}^R-\underline{G}^A) \nonumber \\ &&\times\left(\tanh(\frac{E'}{2k_bT})+\coth(\frac{\omega}{2k_bT})\right),
\end{eqnarray}
with $\omega=E-E'$.
Similarly, the nonquilibrium collision integral in the second term on the right hand side of  Eq.~(\ref{qbe-1}) can be written up to the first order of $\delta \underline{G}^K$ as
\begin{eqnarray}\label{self-energy-2}
G^R\delta\Sigma^K-\delta\Sigma^K G^A=\delta \underline{G}^K\circ D^K (G^R-G^A)\, .
\end{eqnarray}
Substituting Eq.~(\ref{self-energy-1-1}),(\ref{self-energy-1-2}),(\ref{self-energy-2}) into the right hand side of Eq.~(\ref{qbe-1}), the non-equilibrium e-e collision integral has the form
\begin{eqnarray}\label{self-energy-3}
&&I_{ee}(\delta G^K,\delta\underline{G}^K)=I_{ee}(\delta G^K)+\underline{I}_{ee}(\delta\underline{G}^K),
\end{eqnarray}
where
\begin{eqnarray}\label{self-energy-3-1}
I_{ee}(\delta G^K)&=& \left(\underline{G}^K_0\circ (D^R-D^A)+(\underline{G}^R-\underline{G}^A)\circ D^K\right)\delta G^K \nonumber \\
&=&i\frac{1}{\tau_{ee}}\delta G^K \, ,
\end{eqnarray}
and
\begin{eqnarray}\label{self-energy-3-2}
\underline{I}_{ee}(\delta\underline{G}^K)&=&\delta \underline{G}^K\circ \left[(D^R-D^A) G^K_0+D^K(G^R-G^A)\right]\nonumber \\
&=&\delta \underline{G}^K\circ (D^R-D^A)(G^R-G^A)\nonumber \\ && \times\left(\tanh(\frac{E}{2k_bT})+\coth(\frac{\omega}{2k_bT})\right).
\end{eqnarray}
We note that $\underline{I}_{ee}$ has a similar integrand as the one used in  calculating the momentum scattering time of e-e interaction shown in Eq.~\ref{self-energy-1-2}.

Unlike the case of only considering impurity scattering,~\cite{Mishchenko:2004_a,Bernevig:2008_a} 
at finite temperature  the collision integral
 contains  inelastic e-e scatterings which scatter the electron to different energy states and make the collision integral of the right hand side of QKE Eq.~(\ref{qbe-1}) more complicated. Therefore we first give a brief discussion of the e-e interaction in the 2DEG. In the experiments considered the spin splitting energy $\Delta_{so}$ is much smaller than the Fermi energy $\epsilon_F=400$ K, say $\Delta_{so}/\epsilon_F \ll 1$,~\cite{Koralek:2009_a} and therefore the screening of the Coulomb interaction in this SOC system is treated to be the same as in the non-SOC system. The inverse screening length in the two dimensional system has the form \cite{Zheng:1996_a}
\begin{equation}\label{screening-2}
\kappa_d=\frac{2\pi e^2 N_0}{\epsilon_0},
\end{equation}
where $N_0$ is the density of state at the Fermi surface, $e$ is the electron charge and $\epsilon_0$ is the dielectric constant in the vacuum. For the GaAs/AlGaAs, $N_0=\frac{m^*}{\pi  \hbar ^2}$ where $m^*=0.065 \, m_0$  is the effective mass of the electron in the quantum well and $m_0$ is the mass of electron in the vacuum.\cite{Winkler:2003_a} 
In this case, $\kappa_d\approx 3.08\times10^{8} ~\rm{cm}^{-1}$, which is much larger than the Fermi wave length $k_F=\sqrt{2\pi n_0}=2.24\times10^{6}~\rm{cm}^{-1}$, where $n_0=8\times 10^{11} ~\rm{cm}^{-2}$ is the density of electrons \cite{Koralek:2009_a}. Therefore the Coulomb interaction is strongly screened in the 2DEG considered
 in the experiments and we can treat the e-e interaction as  angle independent scattering. On the other hand, the peak of the enhanced-life time of the SHM happens around $75~\rm{K}$  which is much smaller than the Fermi temperature $400~\rm{K}$. \cite{Koralek:2009_a} The non-equilibrium electrons distribute around the Fermi surface within the energy range of $k_BT$ ,where $k_B$ is the Boltzmann constant and $T$ here is the system temperature without confusing it with the time variant $T$ in Eq.~(\ref{qbe-1}). Therefore the non-equilibrium electrons around the turning point $75~\rm{K}$ is in the regime $\epsilon_F \gg k_B T \gg |\epsilon_k-\epsilon_F|$, where $\epsilon_F$ is the Fermi energy and $\epsilon_k$ is the electron energy with  momentum $k$. The e-e scattering time in this regime is estimated theoretically as \cite{Zheng:1996_a}
\begin{eqnarray}\label{screening-3}
\frac{1}{\tau_{ee}}=\frac{\pi \epsilon_F}{8\hbar}\left(\frac{k_B T}{\epsilon_F}\right)^2\ln\frac{\epsilon_F}{k_B T}
\,\,.
\end{eqnarray}
 It is noted that the e-e scattering time in Eq.~(\ref{screening-3}) is independent on the energy of the electrons and there is only one parameter, $\epsilon_F$, that we need from the experimental data to estimate the e-e scattering time.
\begin{figure}
\centering
\begin{tabular}{l}
\includegraphics[width=0.8\columnwidth]{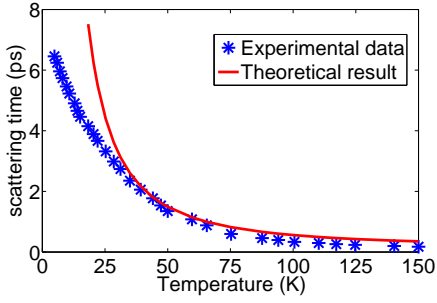}
\end{tabular}
\caption{$\tau_p$ vs T. The blue stars are the momentum scattering times extracted from of $D_s$ vs T in the supplementary material in Ref.~\onlinecite{Koralek:2009_a} by using the relation $\tau=2 D_s/v_F^2$ where $v_F=\frac{\hbar k_F}{m^*}$ is the Fermi velocity and is estimated to be $4.32\times 10^5 ~\rm{m/s}$. The red line is the theoretical estimate of $\tau_{e-e}$ based on the theory of the Ref.~\onlinecite{Zheng:1996_a}.}
\label{life-T-1}
\end{figure} 
In Fig.~\ref{life-T-1}, we compare the momentum scattering time extracted from the experimental data in the supplementary material in Ref.~\onlinecite{Koralek:2009_a} and the theoretically estimate based on Eq.~(\ref{screening-3}). They match very well when the temperature is above $30~\rm{K}$. The mismatch below $30~\rm{K}$ is because the non-equilibrium electrons excited by the optical field are beyond the energy range of $k_B T$ around the Fermi surface in the case of $T<30~\rm{K}$ and Eq.~(\ref{screening-3}) fails in this temperature regime. However, in a wide temperature regime, from $30~\rm{K}$ to $150~\rm{K}$, Fig.~\ref{life-T-1} indicates that the momentum scattering time felt by the spin relaxation is dominated by the e-e scattering time which is consistent with the spin Coulomb drag \cite{Damico:2000_a,Damico:2003_a} observed in the Ref.~\onlinecite{Weber:2005_a}. Therefore, in the following discussion at finite $T> 30$ K, we can safely neglect the impurity scattering in our theory even though it can 
be trivially in corporate when extending the results to lower temperatures.

%It is worth noting that the SOCs enter the quantum kinetic equation Eq.~\ref{qbe-1} in two places. One is from kinetic part, the left hand side of of Eq.~\ref{qbe-1}, such as the effective magnetic field $\bm{h}_{so}$ which exerts a torque on the electron spin and dominates the spin precession. The other one is from the Green functions' part. For example retarded (advanced) Green's function Eq. \ref{Green-1} has off diagonal terms whose contribution to the kinetic equation is proportional to $\Delta_{so}/E_f$ where $\Delta_{so}$ is the spin splitting due to SOCs. In the 2DEG system, $\Delta_{so}/E_f\ll 1$, therefore it will be safely neglected in our following discussion.

\subsection{Thermal average quantum kinetic equation}
\label{QKE-thermal-average}
In the equilibrium state, the Keldysh Green\rq{}s function satisfies \cite{Rammer:2007_a}
\begin{eqnarray}\label{GF-3}
\hat{G}^K_0(E,{\bf k})=(\hat{G}^R-\hat{G}^A)\tanh(\frac{E-\epsilon_F}{2k_B T}).
\end{eqnarray}
 When the quasiparticle approximation is valid, the Keldysh Green\rq{}s function in the $E-{\bf k}$ space is still a peak even in the non-equilibrium state and has the form
  \begin{eqnarray}\label{GF-5}
  \hat{G}^K(E,\rm{k};T,\rm{R})=-2\pi i \delta(E-\epsilon_k) \hat{h},
  \end{eqnarray}
  where $\hat{h}({\bf k},R,T)$ is the distribution function, defined as
 \begin{eqnarray}\label{coll-2}
\hat{h}_{\bm{k}}(\bm{R},T)&=&-\int_{-\infty}^{\infty}\frac{dE}{2\pi i} \hat{G}^{K}_{\bm{k},E}(\bm{R},T)\nonumber \\
&=&g_c\sigma_0+g_x\sigma_x+g_y\sigma_y+g_z\sigma_z.
\end{eqnarray}
In the linear response limit, the non-equilibrium distribution function takes the form
\begin{eqnarray}\label{density-1}
\hat{h}_{\rm{k}}(R,T)=-\frac{f'(\epsilon_k)}{N_0}\hat{g}(\theta,R,T),
\end{eqnarray}
where $N_0$ is the density of state, $\theta$ is the angle between $\rm{k}$ and the $x$ axis, and
\begin{eqnarray}\label{density-2}
\hat{g}=\int N_0 d\epsilon_k \hat{h}_{\rm{k}}(R,T)
\end{eqnarray}
is the thermal average distribution function.
We also introduce the density operator
  \begin{eqnarray}\label{GF-6}
  \hat{\rho}(E,{\rm R},T)&=&i\int\frac{dE}{2\pi} \frac{d^2k}{(2\pi)^2} G^K_{k,E}({\rm R},T)\nonumber \\
  &=&\int \frac{d^2 k}{(2\pi)^2} \hat{h}_{\rm{k}}({\rm R,T})\nonumber \\
  &=& \int \frac{d\theta}{2\pi} \int N_0 \hat{h}_{\rm{k}}({\rm R},T) d\epsilon_k \nonumber \\
  %&=& \int \frac{d\theta}{2\pi}\hat{g}(-\int d\epsilon f'(\epsilon_k))\nonumber \\
  &=& \int \frac{d\theta}{2\pi} \hat{g}(\theta,\rm{R},T).
  \end{eqnarray}
Multiplying by $-\frac{N_0}{2 \pi}$ and integrating over $\epsilon_k$  and $E$ on both sides of Eq.~(\ref{qbe-1}), the left hand side takes the form
\begin{eqnarray}\label{GF-7-1}
\partial_T \hat{g}+\bm{\nabla_R}\cdot \{\frac{1}{2}\bm{\overline{\hat{V}}},  \hat{g} \}+i[\overline{\bm{b}}\cdot \bm{\hat{\sigma}},\hat{g}]+\frac{\hat{g}}{\tau_{ee}},
\end{eqnarray}
where
\begin{eqnarray}\label{GF-7-1-1}
\overline{\hat{V}}&=&\int -f'(\epsilon_k) \hat{V}(\rm{k}) d\epsilon_k=\hat{V}(\overline{k},\theta), \nonumber \\
\overline{\bm{b}}&=&\int -f'(\epsilon_k) \bm{b}(\bm{k})) d\epsilon_k=\overline{\bm{b}}(\overline{k},\overline{k^3},\theta),
\end{eqnarray}
and the right hand side takes the form
 \begin{eqnarray}\label{GF-7-2}
 &&-\int N_0 d\epsilon_k \frac{dE}{2\pi i}\underline{I}_{ee}=-\int N_0 d\epsilon_k\frac{dE}{2\pi i}\frac{dE'}{2\pi}\frac{d^2k'}{(2\pi)^2}\times \nonumber \\ &&\delta \underline{G}^K (D^R-D^A)(G^R-G^A) \nonumber \\ &&\times\left(\tanh(\frac{E}{2k_bT})+\coth(\frac{\omega}{2k_bT})\right)\nonumber \\
 &=& -\int \frac{dE'}{2\pi}\frac{d^2k'}{(2\pi)^2}\delta \underline{G}^K \times \int\frac{dE}{2\pi} N_0 d\epsilon_k (D^R-D^A) \nonumber \\ &&\times (G^R-G^A) \left(\tanh(\frac{E}{2k_bT})+\coth(\frac{\omega}{2k_bT})\right)\nonumber \\
 &\approx&\hat{\rho}(R,T) \times \frac{1}{\tau_{ee}}.
\end{eqnarray}
Therefore, the kinetic equation in Eq.~(\ref{qbe-1}) is converted to the equation of the thermal average distribution $\hat{g}(\theta,\bm{R},T)$ and the density function $\hat{\rho}(\rm{R},T)$ which takes the form
\begin{eqnarray}\label{GF-7}
\partial_T \hat{g}+\bm{\nabla_R}\cdot \{\frac{1}{2}\bm{\overline{\hat{V}}},  \hat{g} \}+i[\overline{\bm{b}}\cdot \bm{\hat{\sigma}},\hat{g}]+\frac{\hat{g}}{\tau_{ee}}=\frac{\hat{\rho}(\bm{R},T)}{\tau_{ee}}\,\,.
\end{eqnarray}
Eq.~(\ref{GF-7}) gives the coupled kinetic equation of the thermal average distribution function $\hat{g}$ and the density matrix $\hat{\rho}(\rm{R},T)$. However, the spin-charge dynamic equation normally is the equation of the density $\hat{\rho}$. Therefore it is necessary to convert  Eq.~(\ref{GF-7})  to an equation only containing $\hat{\rho}$. This is not easily done in the presence of SOCs. Mishchenko \emph{et al.}~\cite{Mishchenko:2004_a} treated the gradient term $\nabla \hat{G}^K$, as a perturbation and developed a self consistent method to derive the diffusion equation to  any order of $\nabla \hat{g}$ in principle. However this method is restricted to the Rashba SOC and becomes harder when calculating the higher order gradient terms or considering the linear and cubic Dresselhaus SOC. 

Bernevig \emph{et al.}~\cite{Bernevig:2008_a} pointed out that in the ballistic regime $ql>1$, where $q$ is the spin-charge wave length and $l$ is the mean free path, we have to consider the higher order gradient terms $\nabla \hat{G}^K$. This means that
in the ballistic regime, the spin-charge dynamic equation is not dominated by the second order of the spacial differential operators $\nabla^2$ but one needs to consider the infinite summation over the gradient expansion. However, this
can only be done for the very special case when $\alpha=\beta_1$ and $\beta_3=0$.\cite{Bernevig:2008_a}  
For a generic SOC, it is a daunting task to evaluate the infinite summation. 
Therefore in this work, we abandon the idea of the gradient expansion of $\nabla \hat{G}^K$ and provide a different way to obtain the spin-charge dynamic equation for the general SOCs.

\subsection{Spin-charge density dynamic equation valid in both weak and strong SOC regime }
\label{general-QKE}
The thermal average distribution function and density matrix can be generally written as
\begin{eqnarray}\label{GF-8}
\hat{g}=g_c\sigma_0+g_x\sigma_x+g_y\sigma_y+g_z\sigma_z \, ,\nonumber \\
\hat{\rho}=\rho_c\sigma_0+\rho_x\sigma_x+\rho_y\sigma_y+\rho_z\sigma_z.
\end{eqnarray}
The third term on the left hand side of Eq.~(\ref{GF-7}) has the form
\begin{eqnarray}\label{Basis-1}
[\overline{b}x \sigma_x+\overline{b}_y \sigma_y, \hat{g}]=2i\left((\overline{b}_x g_y-\overline{b}_y g_x)\sigma_z-\overline{b}_x g_z \sigma_y+ \overline{b}_y g_z \sigma_x \right).\nonumber \\
\end{eqnarray}
This  couples the different spin components and generates spin precession.
The second term on the left hand side of Eq.~(\ref{GF-7}) contains the SOC velocity operators $\hat{v}_{so}=\partial H_{so}/ \partial_k$ which gives
\begin{eqnarray}\label{Basis-3}
&&\{\lambda_1 \sigma_y \partial_x+\lambda_2 \sigma_x \partial_y,\hat{g}\}
\nonumber \\ &=&(\lambda_1 \partial_x g_y+\lambda_2 \partial_y g_x)\sigma_0
+\lambda_1\partial_x g_c \sigma_y+\lambda_2 \partial_y g_c \sigma_x.
\end{eqnarray}
This couples the charge and spin  through the finite $\bm{\nabla}\hat{g}$ which indicates non-uniform charge or spin distribution in real space. The other terms of Eq.~(\ref{Basis-1}) do not couple spin or charge components. If we multiply $\sigma_i$ where $i=0,x,y,z$ and calculate the trace, using the fact that $\rm{Tr(\sigma_i\sigma_j)/2=\delta_{ij}}$, Eq.~(\ref{GF-7}) can be rewritten as
\begin{eqnarray}\label{Basis-4}
&&\left( (\partial_T+\frac{1}{\tau}+\frac{\overline{\bm{k}}}{m}\cdot \bm{\nabla}) g_c+\lambda_1 \partial_x g_y+\lambda_2 \partial_y g_x \right)\sigma_0=\frac{\rho_c}{\tau} \sigma_0 ,\nonumber \\
&&\left( (\partial_T+\frac{1}{\tau}+\frac{\overline{\bm{k}}}{m}\cdot \bm{\nabla}) g_x+\lambda_2 \partial_y g_c+2i \overline{b}_y g_z  \right)\sigma_x=\frac{\rho_x}{\tau} \sigma_x ,\nonumber \\
&&\left( (\partial_T+\frac{1}{\tau}+\frac{\overline{\bm{k}}}{m}\cdot \bm{\nabla}) g_y+\lambda_1 \partial_x g_c-2i \overline{b}_x g_z  \right)\sigma_y=\frac{\rho_y}{\tau} \sigma_y ,\nonumber \\
&&\left( (\partial_T+\frac{1}{\tau}+\frac{\overline{\bm{k}}}{m}\cdot \bm{\nabla}) g_z+2i(\overline{b}_x g_y-\overline{b}_y g_x) \right)\sigma_z=\frac{\rho_z}{\tau} \sigma_z ,\nonumber \\
\end{eqnarray}
where $\overline{\bm{k}}$ is the thermal average of the momentum.
Above we have multiplied again by the corresponding matrix to remind ourselves of which component belongs to
which. Hence, we can obtain the $4\times 4$ kinetic equation of the coefficients $g_{c(x,y,z)}$ and $\rho_{c(x,y,z)}$ which takes the form
 \begin{eqnarray}\label{GF-9}
\hat{K} \left(\begin{array}{c} g_c\\ g_x\\ g_y\\ g_z \end{array}\right)=\left(\begin{array}{c} \rho_c \\ \rho_x \\ \rho_y \\ \rho_z \end{array} \right),
%\nonumber \\
%\left(\begin{array}{c} g_c\\ g_x\\ g_y\\ g_z \end{array}\right)=\hat{K}^{-1}\left(\begin{array}{c} \rho_c \\ %\rho_x \\ \rho_y \\ \rho_z \end{array} \right),
 \end{eqnarray}
 where
 \begin{eqnarray}\label{GF-10}
 &&\hat{K}=\left(
\begin{array}{cccc}
 \Tilde{\Omega}  &-i\lambda_2 q_y\tau &i\lambda_1 q_x\tau & 0 \\
-i\lambda_2 q_y\tau & \Tilde{\Omega}  & 0 & -2 \overline{b}_y\tau  \\
i\lambda_1 q_x \tau  & 0 & \Tilde{\Omega}  & 2 \overline{b}_x\tau \\
 0 & 2 \overline{b}_y\tau  & -2 \overline{b}_x\tau  & \Tilde{\Omega}
\end{array}
\right)\, ,\nonumber \\
\end{eqnarray}
 $\Tilde{\Omega}=1-i\omega \tau+\bm{q}\cdot\bm{v}\tau$, and $\theta$ is angle between $\bm{k}$ and the
 $x$-axis. Here we have Fourier transformed $\partial_T$ and $\partial_{x(y)}$ to $-i\omega$ and $iq_{x(y)}$ which are the frequency and wave-vector of the spin polarization wave in the experiments.
%We also make the approximation that
%   \begin{eqnarray}\label{GF-10-1}
%   G^R \hat{\rho}-\hat{\rho}G^A \approx (G^R-G^A) \hat{\rho}
%   \end{eqnarray} because the non-commute part $G^{R(A)}$ and $\hat{\rho}$ give the correction at the order of $\Delta_{so}/\epsilon_f$ which is very small and can be neglected.
%For convenience, throughout this paper we will express frequency in units of $1/\tau$ and wave vector in units of $1/l$, unless otherwise stated.
 To further obtain the spin dynamic equation, we simply multiply $\hat{K}^{-1}$ on both sides of Eq.~(\ref{GF-7}) and integrate out the angle $\theta$. The inverse of $\hat{K}$ is easily  obtained for the generic SOCs because it is just the inverse of a $4 \times 4$ matrix. This  provides a way to derive the spin dynamic equation in the presence of the general SOCs.

 Before  discussing  the spin dynamic equation, we would like to emphasize  two advantages of our method. First, the inverse of the matrix $\hat{K}$ is equivalent to the infinite summation of the gradient expansion but much easier to be calculated exactly. In the Sec.~\ref{SSOC}, it is shown that in the case of $\alpha=\beta_1$ and $\beta_3=0$, we obtain the same spin dynamics modes to those obtained by the infinite summation of the gradient expansion of $\nabla \hat{G}^K$.~\cite{Bernevig:2008_a} Second, we need not to guess the form of the non-equilibrium distribution function $\hat{g}$ before we obtain the spin dynamics equation. The non-equilibrium distribution function is often  expanded on spherical harmonics (Eq.~(7.79) in   Ref.\onlinecite{Rammer:2007_a}) before we solve the kinetic equation. Normally in the system without SOCs, it is enough to stop the expansion on the first order of the spherical harmonics. However, in the presence of the SOCs, especially considering the cubic Dresselhaus SOC, the distribution function may contain the spherical harmonics up to third order. Therefore it is very difficult and complicated to guess the correct form of the non-equilibrium distribution function by expanding it to the right order in spherical harmonics. Our method does not have this difficulty and is approximate to considering 
 all possible spherical harmonics in the non-equilibrium distribution function.

To simplify our discussion of the spin dynamic equation, we consider the spin wave vector only along the $x$-direction and take $q_y=0$. Because $q\ll k_F$, the spin-charge coupling terms $i\lambda_{1(2)}q_{x(y)}\tau$ are much smaller than the spin-spin coupling terms $i\lambda_{1(2)}k\tau$ in Eq.~(\ref{GF-10}) and we neglect the spin-charge coupling and only focus on the spin-space of Eq.~(\ref{GF-10}). In this case, using the definition
Eq.~(\ref{GF-6}) the spin-charge dynamic equation of the density coefficient $\rho_{c(x,y,z)}$ can be obtained as
 \begin{eqnarray}\label{GF-12}
  \int \frac{d\theta}{2\pi}  \left(\begin{array}{c} g_x\\ g_y\\ g_z \end{array}\right)=\left(\begin{array}{c} \rho_x \\ \rho_y \\ \rho_z \end{array} \right)=\hat{D} \left(\begin{array}{c}  \rho_x \\ \rho_y \\ \rho_z \end{array} \right),
 \end{eqnarray}
where $\hat{D}=\int \frac{d\theta}{2\pi}\hat{K}^{-1}_s$ and $\hat{K}_s^{-1}$ takes the form
%\begin{widetext}
\begin{eqnarray}\label{GF-13}
\hat{K}_s^{-1}=\frac{\left(
\begin{array}{ccc}
 \Tilde{\Omega}^2+4 \overline{b}^2_x \tau ^2  & 4 \overline{b}_x \overline{b}_y \tau ^2  & 2 \overline{b}_y \tau  \Tilde{\Omega}  \\
  4 \overline{b}_x \overline{b}_y \tau ^2 &\Tilde{\Omega}^2+ 4 \overline{b}^2_y \tau ^2  & -2 \overline{b}_x \tau  \Tilde{\Omega}  \\
 -2 \overline{b}_y \tau  \Tilde{\Omega}  & 2\overline{b}_x \tau  \Tilde{\Omega}   & \Tilde{\Omega}^2
\end{array}
\right)}{\Tilde{\Omega}^3+4 \overline{b}^2 \tau ^2 \Tilde{\Omega}}.
\end{eqnarray}
%\end{widetext}
Eq.~(\ref{GF-12}) describes the spin dynamics in the frequency-momentum space at finite temperature and any strength of the SOC. This is the key result of our theory.

\section{Spin dynamics in the weak SOC regime at finite T}\label{PSH}
In this section, we focus on the spin dynamics in the weak SOC regime which is found 
at  temperatures above $35$ K in the experiments considered.~\cite{Koralek:2009_a}

\subsection{Only DP spin relaxation mechanism}
In the regime where $\lambda_{1(2)} k \tau \ll 1$, $ql\ll 1$, defined as the weak SOC regime,
the spin-charge dynamic equation can be written as
\begin{eqnarray}\label{qbe-3}
&&(-i\tilde{\omega}+\frac{1}{2}(\tilde{q}_x^2+\tilde{q}_y^2))\left(\begin{array}{c} \rho_x \\ \rho_y\\ \rho_z  \end{array} \right)+\hat{D}_{so}\left(\begin{array}{c} \rho_x \\ \rho_y\\ \rho_z  \end{array} \right)=0 \, ,
\end{eqnarray}
where
\begin{eqnarray}
&&\hat{D}_{so}=\nonumber \\
&&\left(\begin{array}{ccc} \tilde{\Omega}_{so}^2(\frac{1}{2}+\tilde{\alpha} \tilde{\beta})& 0 &i(\tilde{\alpha}+\tilde{\beta})\tilde{\Omega}_{so}\tilde{q}_x\\0& \tilde{\Omega}_{so}^2 (\frac{1}{2}-\tilde{\alpha}\tilde{\beta})&i(\tilde{\alpha}-\tilde{\beta})\tilde{\Omega}_{so}\tilde{q}_y\\
-i(\tilde{\alpha}+\tilde{\beta})\tilde{\Omega}_{so}\tilde{q}_x &-i(\tilde{\alpha}-\tilde{\beta})\tilde{\Omega}_{so}\tilde{q}_y&\tilde{\Omega}_{so}^2  \end{array} \right) \,,\nonumber \\
&&\tilde{\alpha}=\frac{\alpha}{\sqrt{\alpha^2+(\beta_1-\beta_3)^2+\beta_3^2}}\,,\nonumber \\
&&\tilde{\beta}=\frac{\beta_1-\beta_3}{\sqrt{\alpha^2+(\beta_1-\beta_3)^2+\beta_3^2}},
\end{eqnarray}
$\tilde{\omega}=\omega\tau$, $\tilde{q}_{x(y)}=q_{x(y)}v\tau$,
$\tilde{\Omega}_{so}=\Omega_{so}\tau$ and $\Omega_{so}=2\sqrt{\alpha^2+(\beta_1-\beta_3)^2+\beta_3^2}k_F$ is the spin precession frequency due to the SOC. The first term on the left hand side of Eq.~(\ref{qbe-3}) is the normal diffusion equation without SOCs. The second term  corresponds to the average torque exerted by the SOC in the momentum space which gives the correction of the diffusion equation due to the generic SOCs. The detailed calculation of the elements of the matrix $\hat{D}$ is shown in App.~\ref{D-elements}.

 Our spin diffusion equation, Eq.~(\ref{qbe-3}), is equivalent to the spin diffusion equation in Ref.~\onlinecite{Stanescu:2007_a} at zero temperature where $k=k_F$. However only considering D-P mechanism is not  enough to explain the temperature dependence of the enhanced-lifetime of the SHM,~\cite{Koralek:2009_a} which first increases with increasing temperature from $5~\rm{K}$ to $75~\rm{K}$ and then decreased with further increasing temperature. Because the density of states is a constant for the 2DEG, the chemical potential shift at finite temperature is of the order of $(\frac{kT}{E_F})^4=0.0039$ when $T=100~\rm{K}$ according to the Sommerfeld expansion. Our numerical calculation gives $E_{\rm{f}}=398.15~\rm{K}$ at $T=100~\rm{K}$ which is consistent to the analytical prediction. Assuming that $\chi(E)=-f'(E)$, at $T=100~\rm{K}$ we have
 \begin{eqnarray}\label{average-1}
 \overline{k}= \int dE -f'(E)k=0.98 k_F \,, \nonumber \\
  \overline{k^3}=\int dE -f'(E)k^3=1.1 k_F^3 \, ,
 \end{eqnarray}
  which proves that the thermal average $\overline{k}$ and $\overline{k^3}$ are not changed too much from $0$ K to $100~\rm{K}$. Therefore the increase of the thermal average cubic Dresselhaus SOC is not sufficiently strong to account for the non-monotonic $T$-dependence of the enhanced-lifetime of the SHM.  Another evidence for this statement is from  Fig. 3 (c) in Ref.~\onlinecite{Koralek:2009_a}. The mobility is reduced to avoid the ballistic crossover and the spin life time is measured in five different temperatures as a function of spin polarized wave vector $q$. It is found that at $q=1.26\times 10^4~\rm{cm}^{-1}$, which is close to the SU(2) points, the spin life time is the maximum at the lowest temperature $5~\rm{K}$ and minimum at the highest temperature $250~\rm{K}$. When $q$ is away from the SU(2) point such as $q=0$ or $q=2.5\times 10^4 ~\rm{cm}^{-1}$, the spin life time is the minimum at the lowest temperature $5~\rm{K}$ and maximum at the highest temperature $250~\rm{K}$. As a result, when $q$ is far away from the SU(2) point, the enhanced-lifetime of spin helix mode matches the description of D-P mechanism that the spin life time is inversely proportional to the momentum scattering time. When $q$ is close to the SU(2) point, the turning point, at which the spin life time changes from increasing to decreasing with increasing  temperature, is lower than $5~\rm{K}$. The turning point at such low temperature can not be the consequence of the thermal average of the cubic Dresselhaus SOC.~\cite{Koralek:2009_a}

\subsection{EY mechanism in the 2DEG with the e-e interaction}

 Because D-P mechanism is not sufficient to account for the $T$-dependence of the SHM life time, we have to consider the E-Y mechanism which is proportional to the momentum scattering time,  opposite to the case of D-P  mechanism. There are two processes involved: that of Elliott \cite{Elliott:1954_a} and that of Yafet \cite{Yafet:1963_a}. In the Elliott process, the scattering potential is spin independent. The spin flip is due to the SOC on the Bloch state, say the admixture of the Pauli up and down spins through the coupling between the conduction and valence bands. 
 %Therefore, the Elliott mechanism has the same origin of the Rashba SOC in the conduction band which is also from the coupling between conduction and valence bands.
 In the Yafet process spin flips come directly from the scattering potential with the well known form $\frac{\hbar}{4m_0c^2}(\bm{\nabla} V\times \bm{P}) \cdot \bm{\sigma}$ so that the potential alone couples opposite spins. In the GaAs/AlGaAs 2DEG, as shown in App. \ref{EY}, the Elliott mechanism is much larger than the Yafet mechanism. As a result, in the following discussion, we only consider the Elliott processes. The Elliott mechanism can be derived from the unitary transformation matrix based on the L\"{o}wdin partitioning.\cite{Winkler:2003_a} The coordinate $\bm{r}$ after unitary transformation takes the form \cite{Nozieres:1973_a}
\begin{eqnarray}\label{E-Y-C-1}
&&\bm{r}_{eff}=\bm{r}-\frac{P^2}{3}(\frac{1}{E_0^2}-\frac{1}{(E_0+\Delta_0)^2}) \bm{k}\times \bm{\sigma}=\bm{r}-\gamma \bm{k}\times \bm{\sigma},\nonumber \\
&&\gamma=\frac{P^2}{3}(\frac{1}{E_0^2}-\frac{1}{(E_0+\Delta_0)^2}) \bm{k}\times \bm{\sigma} \, ,
\end{eqnarray}
where $P=\frac{i \hbar^2 }{m_0} <S| \nabla |R>$, $|S\rangle$ is the s-wave like local orbital state, $R=X,Y,Z$ are the p-wave like local orbital states, $E_0$ is the band gap between $\Gamma^-_6$ conduction band and $\Gamma^+_8$ valence band and $\Delta_0$ is the SOC gap between the bands of $\Gamma_7^+$ and $\Gamma_8^+$. \cite{Winkler:2003_a} Any coordinates dependent potential will be modified by the second term in Eq.~(\ref{E-Y-C-1}).
For example, the impurity scattering potential has the form
\begin{eqnarray}\label{E-Y-C-4}
V_{imp}(\bm{r}_{eff})=V_{imp}(\bm{r})-\gamma (\bm{\nabla_{r}}V_{imp}(\bm{r}) \times \bm{k})\cdot\bm{\sigma},
\end{eqnarray} and the Coulomb interaction now has the form
\begin{eqnarray}\label{E-Y-C-2}
V_{e-e}(\bm{r}_{1eff},\bm{r}_{2eff})&=&V_{e-e}(\bm{r_1},\bm{r_2})-\gamma (\bm{\nabla_{r_1}}V_{e-e}\times \bm{k}_1)\cdot \bm{\sigma}\nonumber \\
&&-\gamma(\bm{\nabla_{r_2}}V_{e-e}\times \bm{k}_2)\cdot \bm{\sigma},
\end{eqnarray}
where the subscripts $1$ and $2$ of the space coordinates represent the two interacting electrons respectively. Therefore the spin-orbit coupled scattering potential has the form
\begin{eqnarray}\label{E-Y-C-3}
\hat{V}_{so}(\bm{r})=-\gamma (\bm{\nabla_{r}}V(\bm{r}) \times \bm{k})\cdot\bm{\sigma},
\end{eqnarray}
where $V(\bm{r})=(V_{imp}+V_{e-e})$ is the spin independent momentum scattering potential. The spin relaxation rate due to the E-Y mechanism in the 3D bulk material has the form \cite{Zutic:2004_a}
\begin{eqnarray}\label{E-Y-C-5}
\frac{1}{\tau_{EY}(\epsilon_k)}=A \left(\frac{\Delta_0}{E_0+\Delta_0}\right)^2 \left(\frac{\epsilon_k}{E_0}\right)^2 \frac{1}{\tau(\epsilon_k)},
\end{eqnarray}
where the numerical factor $A$ is of the order of $1$ and dependent on the scattering mechanism \cite{Zutic:2004_a} such as e-e interaction \cite{Boguslawski:1980_a} and $\tau(\epsilon_k)$ is the momentum scattering time at energy $\epsilon_k$. Because the non-equilibrium electrons distribute around the Fermi surface within the energy range $k_B T$, the e-e scattering time is independent on the energy $\epsilon_k$ according to Eq.~(\ref{screening-3}) and is labeled as $\tau$. In the 2DEG, the momentum $k_z$ is quantized. In this case, although the average of $\langle k_z \rangle=0$, $\langle k_z^2 \rangle \approx (\frac{\pi}{d})^2$ which gives the linear Dresselhaus SOC.\cite{Koralek:2009_a} For the same reason, when considering E-Y mechanism, we also assume $\langle k_z^2 \rangle \approx (\frac{\pi}{d})^2$ where $d$ is the width of the quantum well. Therefore, the E-Y mechanism in the 2DEG can be written as
 \begin{eqnarray}\label{E-Y-1}
 \frac{1}{\tau_{EY,x}}=\frac{1}{\tau_{EY,y}}&=&A(1+\frac{2\langle k_z^2 \rangle}{k_F^2})\left(\frac{\Delta_0}{E_0+\Delta_0}\right)^2 \left(\frac{\epsilon_k}{E_0}\right)^2 \frac{1}{\tau},\nonumber \\
 \frac{1}{\tau_{EY,z}}&=&A\frac{4\langle k_z^2 \rangle}{k_F^2}\left(\frac{\Delta_0}{E_0+\Delta_0}\right)^2 \left(\frac{\epsilon_k}{E_0}\right)^2 \frac{1}{\tau},
 \end{eqnarray}
 where $\tau_{\rm{EY},x(y,z)}$ are the spin relaxation time for the spin polarization along $x$, $y$, $z$ direction respectively due to the E-Y mechanism.
 Appendix~\ref{EY} gives the detailed derivation of the E-Y mechanism in the systems considered here.

 \subsection{Temperature dependence of the spin relaxation modes}
\label{T-dep-of-spin-relaxation}
 In this sub-section, we show that the temperature dependence of the spin relaxation modes, especially the enhanced-lifetime of the SHM, is the consequence of the competition between the D-P and E-Y mechanisms.
 By adding the E-Y mechanism on the Eq.~(\ref{qbe-3}), the spin-charge dynamics containing both D-P and E-Y mechanism takes the form
 \begin{widetext}
\begin{eqnarray}\label{qbe-4}
(-i\tilde{\omega}+\frac{1}{2}(\tilde{q}_x^2+\tilde{q}_y^2))\left(\begin{array}{c} n_x \\n_y\\n_z  \end{array} \right)+\left(\begin{array}{ccc} \tilde{\Omega}_{so}^2(\frac{1}{2}+\tilde{\alpha}\tilde{\beta})+\kappa_1^{\shortparallel}& 0 &i(\tilde{\alpha}+\tilde{\beta})\tilde{\Omega}_{so}\tilde{q}_x\\0& \tilde{\Omega}_{so}^2 (\frac{1}{2}-\tilde{\alpha}\tilde{\beta})+\kappa_1^{\shortparallel}&i(\tilde{\alpha}-\tilde{\beta})\tilde{\Omega}_{so}\tilde{q}_y\\
-i(\tilde{\alpha}+\tilde{\beta})\tilde{\Omega}_{so}\tilde{q}_x &-i(\tilde{\alpha}-\tilde{\beta})\tilde{\Omega}_{so}\tilde{q}_y&\tilde{\Omega}_{so}^2+\kappa_1^{\perp}  \end{array} \right)\left(\begin{array}{c} n_x \\n_y\\n_z  \end{array}\right)=0.
\end{eqnarray}
\end{widetext}
where
\begin{eqnarray}\label{kappa-1}
\kappa_1^{\shortparallel}&=&A(1+\frac{2\langle k_z^2 \rangle}{k_F^2})\left(\frac{\Delta_0}{E_0+\Delta_0}\right)^2 \left(\frac{\epsilon_k}{E_0}\right)^2,\nonumber \\
\kappa_1^{\perp}&=&A\frac{4\langle k_z^2 \rangle}{k_F^2}\left(\frac{\Delta_0}{E_0+\Delta_0}\right)^2 \left(\frac{\epsilon_k}{E_0}\right)^2 .
\end{eqnarray}
Here, according to Eq.~(\ref{average-1}), we take the thermal average of the momentum $\overline{k}=k_F$.
To simplify our discussion, we assume that the spin polarization is uniform in $y$ direction and non-uniform in $x$-direction with finite $\tilde{q}_x$. In this case the spin polarization along $y$-direction is not coupled to other two components of spin polarization. Based on the Eq. ~(\ref{qbe-4}), the eigen-mode of the spin polarization along $y$-direction, has the form
\begin{eqnarray}\label{qbe-4-sol-1}
i\tilde{\omega}_y&=&\frac{1}{2}\tilde{q}_x^2+\tilde{\Omega}_{so}^2 (\frac{1}{2}-\tilde{\alpha}\tilde{\beta})+\kappa_1^{\shortparallel},
\end{eqnarray}
where $\tilde{\omega}_y$ is the normalized eigen-frequency of spin polarization along $y$-direction. At the $SU(2)$ symmetric point,~\cite{Bernevig:2006_a} where $\alpha=\beta_1$ and $\beta_3=0$, $(1/2-\tilde{\alpha}\tilde{\beta})=0$ and the effect of the D-P mechanism on the spin polarization along $y$-direction is zero. This is consistent to the fact that the effective magnetic field due to SOC in this case is always along the $y$-direction.

\begin{figure}
\centering
\begin{tabular}{l}
\includegraphics[width=0.8\columnwidth]{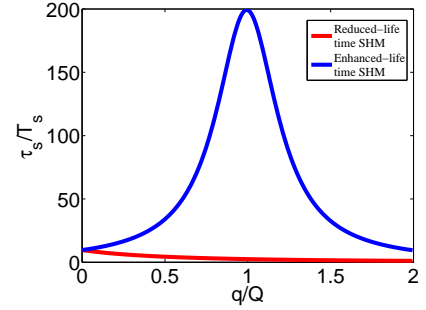}
\end{tabular}
\caption{The reduced-lifetime and enhanced-lifetime of the SHMs. The $x$ axis is the spin wave vector normalized by $Q=4m_{eff}\alpha$ and the $y$ axis is the spin life time normalized by the spin precession period $T_s=\frac{2\pi}{\Omega_{so}}$. $\Omega_{so}\tau=0.1$ is taken to satisfy the weak SOC condition and $\beta_3/\beta_1=0.16$ from the experimental data.\cite{Koralek:2009_a} The maximal enhanced-lifetime is still very close to $Q$ although the cubic Dresselhaus SOC is nonzero.}
\label{2mods}
\end{figure}

Next we focus on the other two spin dynamic modes which have the form
\begin{eqnarray}\label{qbe-4-sol-5}
i\tilde{\omega}_{\pm}&=&\kappa_1^s+\frac{1}{2}\tilde{q}_x^2+\tilde{\Omega}_{so}^2(\frac{3}{4}+\frac{\tilde{\alpha}\tilde{\beta}}
{2})\nonumber \\
&&\pm\sqrt{\left(\tilde{\Omega}_{so}^2(\frac{\tilde{\alpha}\tilde{\beta}}{2}-\frac{1}{4})+\kappa_1^a\right)^2
+(\tilde{\alpha}+\tilde{\beta})^2\tilde{\Omega}_{so}^2 \tilde{q}_x^2},\nonumber \\
\kappa_1^s&=&\frac{\kappa_1^{\shortparallel}+\kappa_1^{\perp}}{2} \ \ \rm{and} \ \ \kappa_1^a=\frac{\kappa_1^{\shortparallel}-\kappa_1^{\perp}}{2}
\end{eqnarray}
  where $\tilde{\omega}_{\pm}$ are the normalized eigen frequencies of the reduced- and enhanced-lifetime of the two SHMs \cite{Koralek:2009_a} respectively. The maximum enhanced-lifetime of the SHM happens when $\alpha=\beta_1$, $\beta_3=0$ and $q=4m_{eff}\alpha=Q$.  In this case, we have $\tilde{\alpha}=\tilde{\beta}=\alpha/\sqrt{\alpha^2+\beta_1^2}=\frac{\sqrt{2}}{2}$ and $\tilde{q}=4m_{eff}\alpha v_f \tau=\sqrt{2}\Omega_{so}\tau=\sqrt{2}\tilde{\Omega}_{so}$. However, in the material such as GaAs, the cubic Dresselhaus SOC is the consequence of bulk inversion asymmetry (BIA) and is inevitable. When considering the cubic Dresselhaus SOC, the maximus spin life time is observed at $\beta_1-\beta_3=\alpha$ \cite{Koralek:2009_a} which is consistent to the theoretical prediction at zero temperature.~\cite{Stanescu:2007_a} When $\beta_3 \ll \alpha$, the maximum enhanced-lifetime of the SHM is still at $Q=4m_{eff}\alpha$ shown in Fig.~\ref{2mods}.
As a result, the Rashba SOC strength can be detected by the relation $\alpha=Q/4m_{eff}$ which gives 
\begin{eqnarray}\label{SOC-1}
\Omega_{so}&=&2\sqrt{\alpha^2+(\beta_1-\beta_3)^2+\beta_3^2}k_F\nonumber \\
&\approx& 2\sqrt{2}\alpha k_F=\frac{Q k_F}{\sqrt{2}m_{eff}}.
\end{eqnarray}

Taking $q_x=4m_{eff}\alpha$ and substituting the above relation to Eq.~(\ref{qbe-4-sol-1}),
we obtain the reduced- and enhanced-lifetime of the SHMs as
\begin{eqnarray}\label{qbe-4-sol-3}
i\omega_{\pm}&=&\frac{\kappa_1^s}{\tau}+\kappa_2^{\pm} \tau,\\
\label{qbe-4-sol-4}
\kappa_2^{\pm}&=&\Omega_{so}^2\left(\frac{7}{4}+\frac{\tilde{\alpha}^2}{2}\pm\sqrt{\left((\frac{\tilde{\alpha}^2}{2}-\frac{1}{4})+
\frac{\kappa_1^a}{\tilde{\Omega}_{so}^2}\right)^2+8 \tilde{\alpha}^2}\right)\, ,\nonumber \\
\end{eqnarray}
where $\pm$ is corresponding to the reduced- and enhanced-lifetime of the SHMs respectively.

We  plot the enhanced-lifetime of the spin helix mode in Fig.~\ref{life-T} and compare our theoretical calculation with Fig.  3 (a) in the Ref.~\onlinecite{Koralek:2009_a}. The system we are considering is the GaAs/AlGaAs quantum well in Ref.~\onlinecite{Koralek:2009_a}. Its width is $11\rm{nm}$ which gives $\beta_3/\beta_1=\langle k_F^2 \rangle/ 4\langle k_z^2 \rangle=0.16$.\cite{Koralek:2009_a} Therefore, in this system, $\beta_1-\beta_3=\alpha \gg \beta_3$ and $\Omega_{so}$ is estimated to be $0.356~\rm{THz}$ based on the Eq.~(\ref{SOC-1}). The band gap $E_0=1.519~\rm{eV}$, $\Delta_0=0.341~\rm{eV}$. To fit the experimental data, Fig.~\ref{life-T}, the only fitting parameter we choose is $A=4.0$ which is the order of 1 and consistent to the estimate in the Ref. \onlinecite{Zutic:2004_a}.
%\begin{figure}
%\centering
%\begin{tabular}{l}
%\includegraphics[width=0.8\columnwidth]{life-T-2.eps}
%\end{tabular}
%\caption{The green stars are the enhanced life time of the spin helix modes extracted from the Ref.~\onlinecite{Koralek:2009_a}. The blue dash line is our theoretical result by substituting $\tau$ in Fig.~\ref{life-T-1} to our spin dynamic equation.}
%\label{life-T}
%\end{figure}

Based on the Eq.~(\ref{qbe-4-sol-3}), both the reduced- and the enhanced-lifetime of the SHMs have a maximal spin life time as a result of the competition between the D-P and E-Y mechanism. The critical momentum scattering times $\tau^{\pm}_c$ when the spin life time reach its maximum take the value
\begin{eqnarray}\label{taumax-1}
&&\tau_c^+=\sqrt{\frac{\kappa_1^s}{\kappa_+}}=0.029 \rm{ps},\nonumber \\
&&\tau_c^-=\sqrt{\frac{\kappa_1^s}{\kappa_-}}=0.51 \rm{ps}.
\end{eqnarray} This is the result of the competition of E-Y and D-P mechanisms. It is noted that $\tau_+ \ll \tau_-$ which indicates the reduced-lifetime needs much higher temperature than the enhanced-lifetime to reach its maximum. Actually $\tau^+_c=0.029~\rm{ps}$ is smaller than the minimum momentum scattering time $0.04~\rm{ps}$ estimated from the spin life time of the enhanced SHM is $100~\rm{ps}$ at $T=300~\rm{K}$ based on the Eq.~(\ref{qbe-4-sol-4}). This is why the reduced-lifetime of the SHM increase monotonically with increasing $T$.

\section{Spin relaxation in the strong SOC regime}\label{SSOC}
In this section we show that our theory can reproduce the theoretical results of the spin relaxation in the case of $\alpha=\beta_1$, $\beta_3=0$  and in the case of only Rashba or linear Dresselhaus SOC in the strong SOC regime at zero temperature.\cite{Bernevig:2008_a,LiuXin:2011_a}
\subsection{$\alpha=\beta_1$}
The persistent SHM occurs in the case of $\alpha=\beta_1$, $\beta_3=0$ and $q$ is along $x$-direction.
In this case, the elements of the matrix $\hat{D}$, corresponding to the reduced- and enhanced-lifetime of the SHMs, in the spin-charge dynamic Eq.~(\ref{GF-12}) take the form
\begin{eqnarray}\label{PSH-GF-1}
D_{11}&=&D_{33}=\frac{1}{2\sqrt{(1-i\tilde{\omega})^2+(\tilde{q} _x+\tilde{\Omega} _{\text{so}})^2}}\nonumber \\&+&\frac{1}{2\sqrt{(1-i\tilde{\omega})^2+(\tilde{q} _x-\tilde{\Omega} _{\text{so}})^2}} \, ,\nonumber \\
D_{13}&=&-D_{31}= \frac{i}{2\sqrt{(1-i\tilde{\omega})^2+(\tilde{q} _x+\tilde{\Omega} _{\text{so}})^2}}\nonumber \\&-&\frac{i}{2\sqrt{(1-i\tilde{\omega})^2+(\tilde{q} _x-\tilde{\Omega} _{\text{so}})^2}},
\end{eqnarray}
where $\tilde{\Omega}_{so}=2\lambda_1 k_F \tau$.
The eigenmodes satisfy the equation
\begin{eqnarray}\label{PSH-GF-2}
(1-D_{11})^2+D_{13}^2=\frac{\left(-1+\sqrt{(1-i\tilde{\omega})^2+\left(\tilde{q} _x-\tilde{\Omega} _{\text{so}}\right){}^2}\right) }{\sqrt{(1-i\tilde{\omega})^2+\left(\tilde{q} _x-\tilde{\Omega} _{\text{so}}\right){}^2} }\nonumber \\
\times \frac{\left(-1+\sqrt{(1-i\tilde{\omega})^2+\left(\tilde{q} _x+\tilde{\Omega} _{\text{so}}\right){}^2}\right)}{\sqrt{(1-i\tilde{\omega})^2+\left(\tilde{q} _x+\tilde{\Omega} _{\text{so}}\right){}^2}}=0,\nonumber \\
\end{eqnarray}
which gives the spin relaxation eigenmodes as
\begin{eqnarray}\label{PSH-GF-2}
i\tilde{\omega}=1\pm\sqrt{1-(\Tilde{q}_x \pm\tilde{\Omega}_{so})^2},
\end{eqnarray}
and is consistent to the result in Ref. \onlinecite{Bernevig:2008_a}.

\subsection{Only Rashba or linear Dresselhaus SOC}

In the presence of Rashba or linear Dresselhaus SOC, we only discuss spin relaxation of the uniform spin polarization because, as far as we are aware of, there is no analytical form of the spin relaxation eigenmodes for the non-uniform spin polarization in the presence of only  Rashba or linear Dresselhaus SOC. For the uniform spin polarization, the off diagonal elements of the dynamic matrix $\hat{D}$ in  Eq.~(\ref{GF-12}) are zero because the angle average of the effective magnetic field $\bm{b}$ is zero. Therefore, we only need to focus on the diagonal terms which have the form
\begin{eqnarray}\label{a-b-1}
D_{11}=D_{22}&=&\frac{\tilde{\Omega}^2+\tilde{\Omega}_{so}^2/2}
{\tilde{\Omega}(\tilde{\Omega}^2+\tilde{\Omega}_{so}^2)},\nonumber \\
D_{33}&=&\frac{\tilde{\Omega}}{\tilde{\Omega}^2+\tilde{\Omega}_{so}^2}.
\end{eqnarray}
The spin relaxation mode of the spin polarization along the $z$-direction satisfies the equation $1-D_{33}=0$ and is solved to have the form
\begin{eqnarray}\label{a-b-2}
i\tilde{\omega}_z=\frac{1}{2}\left(1-\sqrt{1-4\tilde{\Omega}_{so}^2}\right),
\end{eqnarray}
which is identical to our previous theoretical result \cite{LiuXin:2011_a} in the zero temperature limit which has been shown to be consistent  with the experimental observation.\cite{Leyland:2007_a}
\section{Conclusion}

We have developed a consistent microscopic approach to explore the spin dynamics in the presence of SOC
of arbitrary strength, impurity scattering and e-e interaction at finite temperature. The e-e interactions are introduced by using the 
non-equilibrium Green's function formalism. To create a full understanding of the spin-dynamics we incorporate both the DP and EY 
mechanism. 
We have shown that near the SU(2) symmetry point,
because the D-P mechanism is suppressed for the SOC enhanced SHM, the E-Y mechanism  dominates in the high temperature for this mode. By choosing the reasonable parameter of the E-Y mechanism, we show that our theory of enhanced-lifetime of the SHM matches the experimental data quantitatively. Our theory is also shown to be able to recover the several previous theoretical results, which are only valid in the non-interacting system and most focus on the zero temperature limit.

We acknowledge Chia-Ren Hu, Artem Abanov, Alexander Finkel'stein and Valery Pokrovsky for very helpful discussion and support from DMR-1105512, ONR-N000141110780, NHARP, and NSF-MRSEC DMR-0820414.

%\subsection{e-e interaction in the two dimensional electron gas}
%In this section, we discuss the e-e interaction in the 2DEG. For simplify the calculation, we neglect the SOCs whose contribution to the e-e interaction is proportional to the $(\Delta_{so}/E_f)^2$ which is very small in our case.

\appendix
\begin{widetext}

\section{Elliott-Yafet mechanism}\label{EY}

In this appendix, we discuss more extensively the spin relaxation due to the Elliott-Yafet \cite{Elliott:1954_a,Yafet:1963_a} mechanism in the \uppercase\expandafter{\romannumeral3}-\uppercase\expandafter{\romannumeral5} semiconductor quantum well. There are two processes involved: the Elliott process and \cite{Elliott:1954_a} and the Yafet process. \cite{Yafet:1963_a}. In the Yafet process, the spin flip is due to the intrinsic spin-orbit coupling of the scattering potential which has the well know form
 \begin{eqnarray}\label{Yafet}
 H_{so}^{intri}=-\frac{\hbar^2}{4m_0^2c^2}\bm{\sigma}\cdot(\bm{k}\times \bm{\nabla}V),
 \end{eqnarray}
 where $m_0$ is the electron mass in the free space and $V$ is the scattering potential which can be impurity, e-ph and e-e scattering. Here to estimate the strength of the Yafet process, we assume $V(\bm{r})$ is a spherical potential which is independent on the direction.
In this case, the SOC Hamiltonian is Eq.~(\ref{Yafet}) is simplified to be
  \begin{eqnarray}\label{Yafet-2}
  H_{so}^{intri}&=&\frac{\hbar^2}{4m_0^2c^2}\bm{\sigma}(\bm{k}\times \bm{e_r}\partial_r V(r)=-\frac{\hbar^2}{4m_0^2c^2}\frac{\partial_r V(r)}{r}(\bm{r}\times \bm{k})\cdot \bm{\sigma}=-\frac{\hbar^2}{4m_0^2c^2}\frac{\partial_r V(r)}{r}(\bm{L}\cdot \bm{\sigma})\nonumber \\
  &=&-\frac{\hbar^2}{4m_0^2c^2}\frac{\partial_r V(r)}{r}\frac{j(j+1)-l(l+1)-s(s+1)}{2},
  \end{eqnarray}
  where $\bm{L}$ is the orbital angular momentum, $j(l,s)$ is the quantum number of the total angular momentum (orbital angular momentum, spin). In the $\Gamma_7^-$ of the III-V semiconductors, $j=3/2$, $l=1$ and $s=1/2$. Therefore the expectation value of $H_{so}^{intri}$ in $\Gamma_7^-$ can be written as
  \begin{eqnarray}\label{Yafet-3}
  \overline{H}_{so}^{\Gamma_8^{v}}=\langle j=\frac{3}{2},l=1|H_{so}|j=\frac{3}{2},l=1\rangle=-\frac{\hbar^2}{4m_0^2c^2}\langle\frac{\partial_r V(r)}{r} \rangle\frac{1}{2}=\frac{\hbar^2}{4m_0^2c^2}\langle\frac{ V(r)}{r^2} \rangle\frac{1}{2}\approx \frac{\hbar^2}{4m_0^2c^2 (a_0/2)^2}\frac{\overline{V}}{2},
  \end{eqnarray}
   where $a_0$ is the lattice constant. Here we assume that $\langle 1/r^2 \rangle =4/a_0^2$. Similar, in the $\Gamma_7^v$ band, $\overline{H}_{so}^{\Gamma_7^v}=-\frac{\hbar^2}{m_0^2c^2 a_0^2}\overline{V}$ and in the $\Gamma_6^c$, $\overline{H}_{so}^{\Gamma_6^c}=0$.

    In the Elliott process, the spin flip is due to the spin-orbit interaction (SOC) on the Bloch state, say the eigenstate of the electron in the conduction band is not the spin eigenstate. However the momentum scattering potential is $\overline{V}$ which is spin independent. From this viewpoint, the Elliott process has the same origin as the Rashba SOC in the \uppercase\expandafter{\romannumeral3}-\uppercase\expandafter{\romannumeral5} semiconductor quantum well.
  
     The difference lies the fact that in the Rashba SOC, the electric field along $z$-direction break the inversion symmetry and the first non-zero term of Rashb SOC is linear dependent on the wave vector $\bm{k}$ of the conduction electrons which make the spin-up and spin-down states with the same momentum $k$  non-degenerate. In the Elliott processes, the scattering potential does not break the inversion symmetry, its nonzero terms are proportional to $k^2$ and spin-up and spin-down electrons with the same momentum are still degenerate.

     Based on the $8\times8$ Kane model, the effective Hamiltonian of the electrons in the conduction band of \uppercase\expandafter{\romannumeral3}-\uppercase\expandafter{\romannumeral5} semiconductors such as GaAs has the form
\begin{eqnarray}\label{Ham-1}
\hat{H}_c&=&\frac{P^2}{3}\left[\frac{2}{E_0}+\frac{1}{E_0+\Delta_0}\right]k^2-                                 \frac{P^2}{3}\left[\frac{1}{E^2_0}-\frac{1}{(E_0+\Delta_0)^2}\right] \bm{\sigma\cdot \nabla}V \times \mathbf{k}\nonumber \\ &+&\frac{P^2}{3}\left[\frac{1}{E^2_0}+\frac{1}{(E_0+\Delta_0)^2}\right]\frac{\hbar^2}{2m_0^2c^2 a_0^2}\bm{\sigma\cdot \nabla}V \times \mathbf{k},
\end{eqnarray}
where $V$ is the spin independent scattering potential, $P=\frac{i \hbar^2 }{m_0} <S| \nabla |R>$, $|S\rangle$ is the s-wave like local orbital state, $R=X,Y,Z$ are the p-wave like local orbital states, $E_0$ is the energy gap between $\Gamma_6^-$ and $\Gamma_8^+$ bands and $\Delta_0$ is the energy gap between $\Gamma_8^+$ and $\Gamma_7^+$ bands \cite{Winkler:2003_a}. The second term in Eq.~(\ref{Ham-1}) is corresponding to the Elliott spin relaxation mechanism and vanish when the SOC gap $\Delta_0=0$. The third term in Eq.~(\ref{Ham-1}) is corresponding to the Yafet spin relaxation mechanism which will not vanish when the SOC gap $\Delta_0=0$. These are consistent to the characters of Elliott and Yafet mechanisms. For the Elliott spin relaxation mechanism, the scattering potential is spin independent, therefore only the SOC gap $\Delta_0$ can provide the necessary SOC to relax spin. In the Yafet mechanism, the scattering potential itself contains SOC which can relax the spin. Therefore the SOC gap $\Delta_0$ is not necessary in this case.

 Now, let us compare the strength of the Elliott and Yafet mechanism in the GaAs 2DQW, where $P=10.493 ~\rm{eV} {\AA}$, and the band gaps $E_0=1.519$ ${\rm eV}$, $\Delta_0=0.341$ eV.~\cite{Winkler:2003_a} Therefore, in the \uppercase\expandafter{\romannumeral3}-\uppercase\expandafter{\romannumeral5} semiconductors, the ration between the Elliot and Yafet mechanism is of the order of
 \begin{eqnarray}\label{E-Y}
 \frac{\left[\frac{1}{E^2_0}-\frac{1}{(E_0+\Delta_0)^2}\right]}{\left[\frac{1}{E^2_0}+\frac{1}{(E_0+\Delta_0)^2}\right]\left[\frac{\hbar^2}{2m_0^2c^2a_0^2}\right]} =4.84 \times 10^5,
 \end{eqnarray} 
 which indicates that the Elliot spin relaxation is much larger than the Yafet spin relaxation in a III-V semiconductor.
 Therefore, in the following derivation, we only focus on the Elliot mechanism. The first term in the Eq.~(\ref{Ham-1}) is the normal kinetic energy and the second term is the SOC terms In 2D quantum well, by applying an electric potential along $z$ direction $V_z=-eEz$, the inversion symmetry is broken and $\langle\bm{\nabla}V_z\rangle$ is nonzero which gives the Rashba SOC. However, in the Elliott process, the potential $V$ is from the impurity scattering, electron-phonon \cite{Elliott:1954_a} and electron-electron interaction \cite{Tamborenea:2003_a}, the average, $\langle \bm{\nabla}V \rangle$, is zero. Therefore the first nonzero term of the Elliott process is proportional to $(\langle \bm{\nabla}V \rangle)^2$. The SOC potential can be written in the momentum space as~\cite{Akkermans:2007_a}
\begin{eqnarray}\label{app-vso}
\hat{V}_{so}(k,k')&=&i \gamma V(\bm{k} \times \bm{k}')\cdot \bm{\sigma}\nonumber \\
&=&i \gamma V\left\{(k_x k'_y-k_y k'_x)\sigma_z+(k_y k_z'-k_z k'_y) \sigma_x+(k_z k'_x-k_x k_z')\sigma_y\right\}=i(V^z\sigma_z+V^x\sigma_x+V^y\sigma_y),
\end{eqnarray}
where $k'$ and $k$ are the momentums of the electrons before and after scattering respectively, $\gamma=\frac{P^2}{3}(\frac{1}{E_0^2} - \frac{1}{(E_0+\Delta_0)^2} )$ and V is the momentum scattering potential.

Before we calculate the Elliott mechanism, let us connect the quantum kinetic equation to the continuous equation by integrating the momentum $\bm{k}$. To simplify our argument, we consider the case where the SOC due to the inversion asymmetry is zero. Eq.~(\ref{qbe-1}) after the integral has the form
\begin{eqnarray}\label{conti-1}
\partial_T\hat{\rho}(E,\bm{R},T)+\bm{\nabla}\cdot \bm{\hat{J}}(E,\bm{R},T)=\int \frac{d^2 k}{(2\pi)^2}\left[(\Sigma^R G^K+\Sigma^KG^A)-(G^R\Sigma^K+G^{K}\Sigma^A)\right].
\end{eqnarray}
Eq.~(\ref{conti-1}) is the continuous equation in the spin-$\frac{1}{2}$ basis. 
Multiplying by  $\sigma_m/2$ on both sides of Eq.~(\ref{conti-1}) and taking the trace over the spin space
we have the traditional continuous equation
 \begin{eqnarray}\label{conti-2}
 \partial_T\rho_m+\bm{\nabla}\cdot \bm{J}_m=\int \frac{d^2 k}{(2\pi)^2}\rm{Tr}\frac{\sigma_m}{2}\left[(\Sigma^R G^K+\Sigma^KG^A)-(G^R\Sigma^K+G^{K}\Sigma^A)\right],
 \end{eqnarray}
 where $m=0,x,y,z$ are corresponding to charge, spin-$x$, spin-$y$ and spin-$z$ respectively.
 If the charge or spin is conserved, the scattering term on the right hand side of Eq.~(\ref{conti-2}) is zero.
Now let us substitute the SOC potential into the scattering term. There are four terms in the collision integral and we first focus on the two terms containing the retarded Green's function or self energy which has the form
\begin{eqnarray}\label{app-vso-2}
\int \frac{d^2 k}{(2\pi)^2}\rm{Tr}\frac{1}{2}\left\{\sigma_m(\hat{\Sigma}^R\hat{G}^K-\hat{G}^R\hat{\Sigma}^K)\right\}&=&\rm{Tr}\left\{\int \frac{d^2k\rq{}}{(2\pi)^{2}}\frac{\sigma_m}{2}\left[\hat{V}_{kk'}G^R(k') \hat{V}_{k'k}G^K(k)-G^{R}(k)\hat{V}_{kk'}G^K(k')\hat{V}_{k'k}\right]\right\}\nonumber\\
&=&\rm{Tr}\left\{\int \frac{d^2k}{(2\pi)^{2}}\frac{d^2k\rq{}}{(2\pi)^{2}}(\frac{\sigma_m}{2}\hat{V}_{kk'}-\hat{V}_{kk'}\frac{\sigma_m}{2})G^R(k') (\hat{V}_{k'k})G^K(k)\right\}\nonumber \\
&+&\rm{Tr}\left\{\int \frac{d^2k}{(2\pi)^{2}}\frac{d^2k\rq{}}{(2\pi)^{2}}\frac{\sigma_m}{2}(G^R(k')\hat{V}_{k'k}\hat{G}^K(k)\hat{V}_{kk'}-G^R(k) \hat{V}_{kk'}\hat{G}^K(k')\hat{V}_{k'k})\right\}.
\end{eqnarray}
The last term on the right hand side of Eq.~(\ref{app-vso-2}) is zero because the symmetry of $k$ and $k'$ in this term. The first term on the right hand side of Eq.~(\ref{app-vso-2}) can be written further as
\begin{eqnarray}\label{app-vso-3}
&&\rm{Tr}\left\{\int \frac{d^2k\rq{}}{(2\pi)^{2}}(\frac{\sigma_m}{2}\hat{V}_{kk'}-\hat{V}_{kk'}\frac{\sigma_m}{2})G^R(k') (\hat{V}_{k'k})G^K(k)\right\}\nonumber \\
&=&\rm{Tr}\left\{\int \frac{d^2k\rq{}}{(2\pi)^{2}}[\frac{\sigma_m}{2},V^{j}\sigma^j]G^R(k') (\hat{V}_{k'k})G^K(k)\right\}= \rm{Tr}\left\{\int \frac{d^2k\rq{}}{(2\pi)^{2}}2i\epsilon_{mjk}V^j_{kk'}\frac{\sigma_k}{2} G^R(k') (\hat{V}_{k'k})G^K(k)\right\}\nonumber \\ &=&\rm{Tr}\left\{\int \frac{d^2k\rq{}}{(2\pi)^{2}}(-4)\epsilon_{mjk}\epsilon_{kjm}V^j_{kk'}V^j_{k'k}\frac{\sigma_m}{2} G^R(k') G^K(k)\right\}=\rm{Tr}\left\{\int \frac{d^2k\rq{}}{(2\pi)^{2}}4(\sum_{j\neq m}V^j_{kk'}V^j_{k'k}) \frac{\sigma_m}{2} G^R(k')G^K(k)\right\},
\end{eqnarray}
For the charge dynamics, we need to take $m=0$. Because $\sigma_0$ commutes with any scattering potential operator,  Eq.~(\ref{app-vso-3}) is always zero which indicates that the charge is always conserved. For the spin dynamics, $m=x,y,z$. Generally speaking, $\sigma_{x(y,z)}$ does not commutes with a spin dependent scattering potential and the Eq.~(\ref{app-vso-3}) is nonzero which indicates that the spin will decay. Note that in the 2DEG, $k_z$ is quantized. Therefore, taking $m=x,y,z$ we have
\begin{eqnarray}\label{app-vso-4}
\rm{Tr}\left\{\int \frac{d^2k\rq{}}{(2\pi)^{2}}4(\sum_{j\neq x}V^j_{kk'}V^j_{k'k}) \frac{\sigma_x}{2} G^R(k')G^K(k)\right\}&=&-i\frac{\gamma^2 k_F^4}{\tau_p}(1+\frac{2k_z^2}{k_F^2})G^K_x=-i A(1+\frac{2\langle k_z^2 \rangle}{k_F^2})\left(\frac{\Delta_0}{E_0+\Delta_0}\right)^2 \left(\frac{\epsilon_k}{E_0}\right)^2 \frac{G^K_x}{2\tau_p},\nonumber \\
\rm{Tr}\left\{\int \frac{d^2k\rq{}}{(2\pi)^{2}}4(\sum_{j\neq y}V^j_{kk'}V^j_{k'k}) \frac{\sigma_y}{2} G^R(k')G^K(k)\right\}&=&-i\frac{\gamma^2 k_F^4}{\tau_p}(1+\frac{2k_z^2}{k_F^2})G^K_y=-i A(1+\frac{2\langle k_z^2 \rangle}{k_F^2})\left(\frac{\Delta_0}{E_0+\Delta_0}\right)^2 \left(\frac{\epsilon_k}{E_0}\right)^2 \frac{G^K_x}{2\tau_p},\nonumber \\
\rm{Tr}\left\{\int \frac{d^2k\rq{}}{(2\pi)^{2}}4(\sum_{j\neq z}V^j_{kk'}V^j_{k'k}) \frac{\sigma_z}{2} G^R(k')G^K(k)\right\}&=&-i\frac{\gamma^2 k_F^4}{\tau_p}\frac{4k_z^2}{k_F^2}G^K_z=-i A\frac{4k_z^2}{k_F^2}\left(\frac{\Delta_0}{E_0+\Delta_0}\right)^2 \left(\frac{\epsilon_k}{E_0}\right)^2 \frac{G^K_z}{2\tau_p},
\end{eqnarray}
where $k_z=\pi/d$ and $d$ is the width of the quantum well and $A$ is the order of unity. The other two terms,  containing the advance Green's function in Eq.~(\ref{conti-2}), gives the same result to the Eq.~(\ref{app-vso-4}). 
The average, $\langle k_z^2 \rangle$, is nonzero which is also the reason why there is a linear Dresselhaus term in the \uppercase\expandafter{\romannumeral3}-\uppercase\expandafter{\romannumeral5} semiconductor quantum well. Therefore, although average $\langle k_z \rangle$ is zero, the components of Elliott mechanism which are proportional to $\langle k_z^2 \rangle$ are still finite. As a result, the Elliott spin relaxation rates have the form
\begin{eqnarray}\label{app-vso-5}
 \frac{1}{\tau_{EY,x}}=\frac{1}{\tau_{EY,y}}&=&A(1+\frac{2\langle k_z^2 \rangle}{k_F^2})\left(\frac{\Delta_0}{E_0+\Delta_0}\right)^2 \left(\frac{\epsilon_k}{E_0}\right)^2 \frac{1}{\tau_p}=\frac{\kappa_1^{\shortparallel}}{\tau_p},\nonumber \\
 \frac{1}{\tau_{EY,z}}&=&A\frac{4\langle k_z^2 \rangle}{k_F^2}\left(\frac{\Delta_0}{E_0+\Delta_0}\right)^2 \left(\frac{\epsilon_k}{E_0}\right)^2 \frac{1}{\tau_p}=\frac{\kappa_1^{\perp}}{\tau_p}.
 \end{eqnarray}
The same result can be obtained from the traditional definition of the spin decay rate due to the admixture of the Pauli spin-up and spin-down in the eigenstates of the conduction electron. From the $8\times 8$ Kane mode we can obtain these eigenstates as
\begin{eqnarray}\label{eigen-1}
\Psi_{ck,\uparrow}&=&|S,\uparrow\rangle+\frac{-1}{\sqrt{2}}\frac{P}{E_0}k_+|\frac{3}{2},\frac{3}{2}\rangle+ \sqrt{\frac{2}{3}}\frac{P}{E_0}k_z|\frac{3}{2},\frac{1}{2}\rangle+\frac{1}{\sqrt{6}}\frac{P}{E_0}k_-|\frac{3}{2},-\frac{1}{2} \rangle\nonumber \\&&+\frac{-1}{\sqrt{3}}\frac{P}{E_0+\Delta_0}k_z|\frac{1}{2},\frac{1}{2}\rangle+\frac{-1}{\sqrt{3}}\frac{P}{E_0+\Delta_0}k_-|\frac{1}{2} ,-\frac{1}{2}\rangle,\nonumber \\
\Psi_{ck,\downarrow}&=&|S,\downarrow\rangle+\frac{-1}{\sqrt{6}}\frac{P}{E_0}k_+|\frac{3}{2},\frac{1}{2}\rangle +\sqrt{\frac{2}{3}}\frac{P}{E_0}k_z|\frac{3}{2},-\frac{1}{2}\rangle+\frac{1}{\sqrt{2}}\frac{P}{E_0}k_{-}|\frac{3}{2},\nonumber \\&& -\frac{3}{2}\rangle+\frac{-1}{\sqrt{3}}\frac{P}{E_0+\Delta_0}k_+|\frac{1}{2},\frac{1}{2}\rangle+\frac{1}{\sqrt{3}} \frac{P}{E_0+\Delta_0}k_z|\frac{1}{2},-\frac{1}{2}\rangle.\nonumber \\
\end{eqnarray}
The transition amplitude between these two states with different wave vector $k$ are proportional to
\begin{eqnarray}\label{collision-6}
\langle \Psi_{ck\rq{},\downarrow}|V(\bm{r})|\Psi_{ck,\uparrow}\rangle&=&V_0\left(-\frac{1}{3}\frac{P^2}{E_0^2}(k_z k_-\rq{}-k_-k_z\rq{})+\frac{1}{3}\frac{P^2}{(E_0+\Delta_0)^2}(k_zk_-\rq{}-k_-k_z\rq{})\right)\nonumber \\
&=&V_0\left(-\frac{1}{3}(k_zk_-\rq{}-k_-k_z\rq{})(\frac{P^2}{E_0^2}-\frac{P^2}{(E_0+\Delta_0)^2})\right)\nonumber \\
&=&-V_0\gamma (k_zk_-\rq{}-k_-k_z\rq{}).
\end{eqnarray}
Therefore the spin decay rate is proportional to
\begin{eqnarray}\label{collision-7}
\frac{1}{\tau_s}\propto V_0^2\gamma^2 (k_zk_-\rq{}-k_-k_z\rq{}) (k_zk_+\rq{}-k_+k_z\rq{})\propto \gamma^2 k_f^4 \frac{1}{\tau_p},
\end{eqnarray}
where $1/\tau_p$ is proportional to the $V_0^2$. Eq.~(\ref{collision-7}) is equivalent to the Elliott spin relaxation time we obtain from the projected Hamiltonian into the conduction band.

 \section{How the e-e interaction vanishes in calculating the charge conductivity}
\label{e-e-c}
 When deriving the quantum kinetic equation, the collision integral contains many kinds of momentum scattering channels; one dominant at finite temperature is the e-e interaction. However as we know that when calculating the conductivity, the e-e interaction should vanish in this case. In this appendix, we show that how the e-e interaction naturally vanish in calculating the conductivity and, therefore, they only affect the spin relaxation channles. 
 To simplify our discussion, we start from the classical Boltzmann equation without SOC
 \begin{eqnarray}\label{2pgf-20}
&&\partial_T f_{k_1} +\frac{1}{2}\{\hat{\mathbf{V}}_k,\cdot \mathbf{\nabla_R}f_{k_1}\}-e\mathbf{E}\cdot \mathbf{\nabla}_k f_{k_1}=\int U_{k_1,k_1'}(f_{k_1}-f_{k_1'})\nonumber \\
&+&\int d^n k_1' k_1,d^n k_2' W(k_1,k_1',k_3,k_4)\left[f(k_1)f(k_2)(1-f(k_1\rq{}))(1-f(k_2\rq{}))-f(k_1\rq{})f(k_2\rq{})(1-f(k_1))(1-f(k_2))\right],\nonumber\\
\end{eqnarray}
where $n$ is the dimension of the system, $f(k)$ is the Fermi distribution function, $U_{k_1,k_1'}$ is the impurity scattering rate and $W(k_1,k_1';k_2,k_2')$ is the e-e scattering rate. If we exchange $f_1$, $f_2$ and $f_1\rq{}$, $f_2\rq{}$, the right hand side of Eq. (\ref{2pgf-20}) is unchanged. This means the electron $2$ satisfies the same Boltzmann equation as electron $1$:
 \begin{eqnarray}\label{2pgf-21}
&&\partial_T f_{k_2} +\frac{1}{2}\{\hat{\mathbf{V}}_k,\cdot \mathbf{\nabla_R}f_{k_2}\}-e\mathbf{E}\cdot \mathbf{\nabla}_k f_{k_2}=\int V(f_{k_2}-f_{k_2'})\nonumber \\
&+&\int d^n k_1' d^n k_2 d^n k_2'W(k_1,k_2,k_3,k_4)\left[f(k_1)f(k_2)(1-f(k_1\rq{}))(1-f(k_2\rq{}))-f(k_1\rq{})f(k_2\rq{})(1-f(k_1))(1-f(k_2))\right],\nonumber\\
\end{eqnarray}
The charge current operator is defined as
\begin{eqnarray}\label{CBZ-3}
J=\int d^n k_1 \left( -e\frac{k_1}{m^*}\right) f(k_1)=\int d^n k_2 \left( -e\frac{k_2}{m^*}\right) f(k_2),
\end{eqnarray}
where $m^*$ is the effective electron mass. Since we are interested in the dc-conductivity, the system is uniform and independent on time. Therefore the first two terms on the left side of Boltzmann equation are zero.
To get the charge current equation from the Boltzmann equation, we multiply the charge current operator on the both sides of Eq.~(\ref{2pgf-20}) and Eq.~(\ref{2pgf-21}) which gives
\begin{eqnarray}\label{CBZ-4}
&& \int dk_1 \frac{e^2 \bm{k}_1}{m}\bm{E}\cdot \bm{\partial_k} f_1 =\int dk_1 dk_1\rq{}\frac{-e\bm{k}_1}{m} U(k_1,k_1\rq{})(f_1-f_1\rq{})\nonumber \\&&+\int dk_1 dk_2 dk_1\rq{} dk_2\rq{} \frac{e\bm{k}_1}{m} W(k_1,k_2,k_1\rq{},k_2\rq{})  \left( f_1 f_2(1-f_1\rq{}) (1-f_2\rq{})-f_1\rq{} f_2\rq{} (1-f_1)(1-f_2)\right).
\end{eqnarray}
\begin{eqnarray}\label{CBZ-5}
&& \int dk_2 \frac{e^2 \bm{k}_2}{m}\bm{E}\cdot \bm{\partial_k} f_2 =\int dk_2 dk_2\rq{}\frac{-e\bm{k}_2}{m} U(k_2,k_2\rq{})(f_2-f_2\rq{})\nonumber \\&&+\int dk_1 dk_2 dk_1\rq{} dk_2\rq{} \frac{e\bm{k}_2}{m} W(k_1,k_2,k_1\rq{},k_2\rq{}) \left( f_1 f_2(1-f_1\rq{}) (1-f_2\rq{})-f_1\rq{} f_2\rq{} (1-f_1)(1-f_2)\right).
\end{eqnarray}

The left hand side of Eq.~(\ref{CBZ-4}) and Eq.~(\ref{CBZ-5}) have the form
\begin{eqnarray}\label{CBZ-6}
 \int dk_{1{2}} \frac{e^2 k_{1(2)}}{m}\bm{E}\cdot \bm{\partial_k} f_{1(2)} =\int dk_{1(2)} \bm{\partial_k}\cdot(\frac{e^2 \bm{k}_{1(2)}}{m} \bm{E} f_{1(2)})-\int dk_{1(2)} (\bm{\partial_k} \frac{e^2 \bm{k}_{1(2)}}{m}) \bm{E} f_{1(2)}=\frac{ne^2}{m},\nonumber \\
 \end{eqnarray}
 where $n$ is the density of the electron and $j=\int dk\frac{ek}{m} f(k)$ is the electric current. The first term on the right hand side of Eq.~(\ref{CBZ-4}) and Eq.~(\ref{CBZ-5}) correspond to the impurity scattering and equal to $\frac{j}{\tau_{imp}}$ where $\tau_{imp}$ is the momentum scattering time due to the impurity. The second term on the right hand side is not easy to calculate because the momentums $k_1$, $k_2$, $k_1'$ and $k_2'$ are not independent but correlated by the fact that the e-e interaction conserve the net momentum, {\it i.e.} ${\bf k}_1+{\bf k}_2={\bf k}_1+{\bf k}_2\rq{}$. However, if we calculate $\frac{1}{2}(Eq.~(\ref{CBZ-4})+Eq~(\ref{CBZ-5}))$, the e-e scattering can be written as
\begin{eqnarray}\label{CBZ-8}
&&\int dk_1 dk_2 dk_1\rq{} dk_2\rq{} \frac{e(\bm{k}_1+\bm{k}_2)}{2m} W(k_1,k_2,k_1)\rq{},k_2\rq{}) \left( f_1 f_2(1-f_1\rq{}) (1-f_2\rq{})-f_1\rq{} f_2\rq{} (1-f_1)(1-f_2)\right)=\nonumber \\
&&\int dk_1 dk_2 dk_1\rq{} W(k_1,k_2;k_1\rq{},k_2\rq{})(\frac{e(\bm{k}_1+\bm{k}_2)}{2m} f_1 f_2(1-f_1\rq{}) (1-f_2\rq{})-\frac{e(\bm{k}_1\rq{}+\bm{k}_2\rq{})}{2m}f_1\rq{} f_2\rq{} (1-f_1)(1-f_2))\nonumber \\
&&=0.
\end{eqnarray}
Therefore the charge current equation, $\frac{1}{2}(Eq.~(\ref{CBZ-4})+Eq.~(\ref{CBZ-5}))$, has the form
\begin{eqnarray}\label{CBZ-7}
\frac{ne^2}{m}=\frac{j}{\tau_{imp}},\nonumber \\
\end{eqnarray}
where e-e scattering term exactly disappear in this current equation as long as the e-e interaction conserves the net momentum which is not sensitive to the form of the e-e scattering rate $W(k_1,k_1';k_2,k_2')$. However, when deriving the density matrix equation, we do not multiply by the current operator and therefore e-e scattering can not be canceled as we did in the current equation. This is reasonable because the density does not care about the net momentum which is the key to get the current equation.
Therefore although the original Boltzmann equation of the distribution function contains the e-e interaction, the electron-electron scattering will disappear when we calculate the charge conductivity. 
 Hence the e-e interaction will not affect the conductivity but can contribute the spin dynamics.

\section{The derivation of the matrix element of the spin dynamic equation}\label{D-elements}
The denominator of Eq.~(\ref{GF-13}) can be expand to
\begin{eqnarray}\label{GF-16}
&&\frac{1}{\Tilde{\Omega} ^2+4 k^2 \tau ^2 \cos^2\theta \lambda _1^2+4 k^2 \tau ^2 \sin^2\theta \lambda _2^2}\nonumber \\&\approx& 1+2i\tilde{\omega}-2i(\tilde{q}_x \cos\theta+\tilde{q}_y \sin\theta)-3\tilde{q}_x^2\cos^2\theta-3\tilde{q}_y^2\cos^2\theta-4 k^2 \tau ^2 \cos^2\theta \lambda _1^2-4 k^2 \tau ^2 \sin^2\theta \lambda _2^2
\end{eqnarray}
where $\tilde{\omega}=\omega\tau$ and $\tilde{q}_{x(y)}=q_{x(y)}v\tau$.
Substituting Eq.~(\ref{GF-16}) to Eq.~(\ref{GF-7}) we have
\begin{eqnarray}\label{GF-17}
D_{13}&=&-D_{31}=\int \frac{d\theta}{2\pi}\frac{2\lambda_1 k \tau \cos\theta}{\Tilde{\Omega} ^2+4 k^2 \tau ^2 \cos^2\theta \lambda _1^2+4 k^2 \tau ^2 \sin^2\theta \lambda _2^2}\nonumber \\
&\approx& -\int \frac{d\theta}{2\pi} 4ik\tau \Tilde{q}_x (\alpha+\beta_1-2\beta_3\cos2\theta)\cos^2\theta=-2i(\alpha+\beta_1-\beta_3)k \tau \Tilde{q}_x \end{eqnarray}\begin{eqnarray}
D_{23}&=&-D_{32}=\int \frac{d\theta}{2\pi}\frac{-2\lambda_2 k \tau \sin\theta}{\Tilde{\Omega} ^2+4 k^2 \tau ^2 \cos^2\theta \lambda _1^2+4 k^2 \tau ^2 \sin^2\theta \lambda _2^2}\nonumber \\
&\approx& \int \frac{d\theta}{2\pi} 4i k \tau \Tilde{q}_y (\beta_1-\alpha_1+\beta_3\cos2\theta)\sin^2\theta =2i(\beta_1-\beta_3-\alpha) k \tau \Tilde{q}_y\end{eqnarray}\begin{eqnarray}
D_{33}&=&\int \frac{d\theta}{2\pi}\frac{\Tilde{\Omega}}{\Tilde{\Omega} ^2
+4 k^2 \tau ^2 \cos^2\theta \lambda _1^2+4 k^2 \tau ^2 \sin^2\theta \lambda _2^2}\nonumber \\ &\approx& \int \frac{d\theta}{2\pi} (1+i\tilde{\omega}-\Tilde{q}_x^2\cos^2\theta-\Tilde{q}_y^2\sin^2\theta-4 k^2 \tau ^2 \cos^2\theta \lambda _1^2-4 k^2 \tau ^2 \sin^2\theta \lambda _2^2) \nonumber \\
&=& 1+i\tilde{\omega}-\frac{1}{2}(\Tilde{q}_x^2+\Tilde{q}_y^2)-4(\alpha^2+(\beta_1-\beta_3)^2+\beta_3^2)k^2\tau^2=1+i\tilde{\omega}-\frac{1}{2}(\Tilde{q}_x^2+\Tilde{q}_y^2)-\Tilde{\Omega}_{so}^2\tau^2 \end{eqnarray}\begin{eqnarray}
D_{11}&=& \int \frac{d\theta}{2\pi} \frac{\Tilde{\Omega}^2+4\lambda_2^2k^2\tau^2\sin^2\theta}{\Tilde{\Omega}(\Tilde{\Omega} ^2+4 k^2 \tau ^2 \cos^2\theta \lambda _1^2+4 k^2 \tau ^2 \sin^2\theta \lambda _2^2)}\nonumber \\
&\approx& D_{33}+\int \frac{d\theta}{2\pi} 4\lambda_2^2k^2\tau^2\sin^2\theta = D_{33}+(\frac{1}{2} \left(\alpha -\beta _1\right){}^2+\left(\alpha -\beta _1\right) \beta _3+\beta _3^2)4k_F^2\tau^2\nonumber \\
&=& 1+i\tilde{\omega}-\frac{1}{2}(\Tilde{q}_x^2+\Tilde{q}_y^2)-2(\alpha^2+(\beta_1-\beta_3)^2+\beta_3^2)k^2\tau^2-
\alpha(\beta_1-\beta_3)\nonumber\\
&=& 1+i\tilde{\omega}-\frac{1}{2}(\Tilde{q}_x^2+\Tilde{q}_y^2)-\frac{1}{2}\Omega_{so}^2\tau^2
4k_F^2\tau^2\alpha(\beta_1-\beta_3)\end{eqnarray}
\begin{eqnarray}
D_{22}&=&\int \frac{d\theta}{2\pi} \frac{\tilde{\Omega}^2+4\lambda_2^2k^2\tau^2\sin^2\theta}{\tilde{\Omega} ^2+4 k^2 \tau ^2 \cos^2\theta \lambda _1^2+4 k^2 \tau ^2 \sin^2\theta \lambda _2^2}\nonumber \\
&\approx& \int \frac{d\theta}{2\pi} (1+i\tilde{\omega}-\Tilde{q}_x^2\cos^2\theta-\Tilde{q}_y^2\sin^2\theta-4 k^2 \tau ^2 \cos^2\theta \lambda _1^2)\nonumber \\
&=&1+i\tilde{\omega}-\frac{1}{2}(\Tilde{q}_x^2+\Tilde{q}_y^2)-2(\alpha^2+(\beta_1-\beta_3)^2
+\beta_3^2)k^2\tau^2+\alpha(\beta_1-\beta_3)
\nonumber \\
&=&1+i\tilde{\omega}-\frac{1}{2}(\Tilde{q}_x^2+\Tilde{q}_y^2)-\frac{1}{2}\Omega_{so}^2\tau^2+4k^2\tau^2\alpha(\beta_1-\beta_3)
\end{eqnarray}
where $\Omega_{so}=2\sqrt{\alpha^2+(\beta_1-\beta_3)^2+\beta_3^2}k$.

\end{widetext}

%%%%%%%%%%%%%%%%%%%%%%%%%%%%%%%%%%%%%%%%%%%%%%%%%%%%%%%%%%%%%%%%%%%%%%%%%%%%%%%%%%%%%%%%%%%%%%%%%%
%\bibliography{spin-charge,TI_references}
%merlin.mbs 2010-03-15 4.21a (PWD, AO, DPC)
%Control: key (0)
%Control: author (8) initials jnrlst
%Control: editor formatted (1) identically to author
%Control: production of article title (-1) disabled
%Control: page (0) single
%Control: year (1) truncated
%Control: production of eprint (0) enabled
%

%%%%%%%%%%%%%%%%%%%%%%%%%%%%%%%%%%%%%%%%%%%%%%%%%%%%%%%%%%%%%%%%%%%%%%%%%%%%%%%%%%%%%%%%%%%%%%%%%%%
\noindent

%%%%%%%%%%%%%%%%%%%%%%%%%%%%%%%%%%%%%%%%%%%%%%%%%%%%%%%%%%%%%%%%%%%%%%%%%%%%%%%%%%%%%%%%%%%%%%%%%%%%%%%%%%%%%%%%%%%%%%%%%%%%

\end{document}